\documentclass[epj,nopacs]{svjour}
\usepackage[utf8]{inputenc}
\usepackage{amsmath}
\usepackage{color}
\usepackage{amsfonts}
\usepackage{amssymb}
\usepackage{graphicx}
\usepackage{hyperref}
\usepackage{bm}

\newcommand{\kv}{{\bd{k}}}
\newcommand{\qv}{{\bd{q}}}
\newcommand{\vfv}[1]{\tilde{\psi}_{#1}}

\newcommand{\pvf}[1]{\tilde{p}_{#1}^c}
\newcommand{\bd}{\bm}

\begin{document}
\title{Collisionless kinetic theory for parametrically pumped magnons}
\date{March 12, 2020}
\author{Viktor Hahn\thanks{e-mail: hahn@itp.uni-frankfurt.de} \and Peter Kopietz}
\institute{Institut f\"{u}r Theoretische Physik, Universit\"{a}t Frankfurt, Max-von-Laue Strasse 1, 60438 Frankfurt, Germany}

\begin{abstract}
{
We discuss collisionless kinetic equations describing the 
non-equilibrium dynamics of magnons
in a ferromagnet exposed to an oscillating microwave field.
Previously, this problem has been treated within the so-called
\textquotedblleft S-theory\textquotedblright\ 
where the 
collision integral in the kinetic equation for the magnon distribution
is  either neglected 
or taken into account  phenomenologically via an effective relaxation time.
However, the possibility of magnon condensation has not been included in S-theory. 
Moreover, 
the  momentum integrations appearing in the magnon self-energies are usually decoupled 
by retaining only the term where the loop momentum is equal to the external momentum.
In this work we critically examine 
the accuracy of these approximations  and develop  the
proper extensions of S-theory. We show that these extensions can significantly modify
the time evolution of the magnon distribution.}
\end{abstract}
\maketitle

\section{Introduction}
Magnons in ordered magnets can be excited by exposing the system to an 
external oscillating microwave field. Due to the periodic modulation of the coupling parameter,
energy can be pumped into the system via parametric resonance.
The theory of parametric resonance in magnon gases has a long history, starting with
pioneering works by  Suhl~\cite{Suhl57} and  by Schl\"{o}mann {\it{et al.}} \cite{Schloemann60}.
A comprehensive theory of parametric resonance in magnon gases was developed in the 1970's by Zakharov, L'vov, and Starobinets \cite{Zakharov70} 
who
derived and solved non-linear kinetic equations 
for the diagonal- and off-diagonal magnon distribution functions
 \begin{subequations}
 \begin{eqnarray}
 n_{\bd{k}} ( t ) &= & \langle a^{\dagger}_{\bd{k}} ( t )
 a_{\bd{k}} ( t ) \rangle,
 \label{eq:nkdef}
  \\
p_{\bd{k}} ( t ) & = & \langle a_{- \bd{k}} ( t ) a_{\bd{k}} ( t ) \rangle ,
 \label{eq:pkdef}
 \end{eqnarray}
 \end{subequations}
within  the time-dependent self-consistent 
Hartree-Fock approximation.
Here $a_\kv(t)$ and $a^\dagger_\kv(t)$ 
are the magnon annihilation and creation operators in the Heisenberg picture
and $\langle \ldots \rangle$ denotes the non-equilibrium statistical average.
The theoretical framework based on the time-dependent self-consistent Hartree-Fock approximation developed by
Zakharov {\it{et al.}} \cite{Zakharov70,Zakharov74}  has been called
\textquotedblleft S-theory\textquotedblright \space and has been quite successful to explain
experiments probing the non-equilibrium magnon dynamics in ferromagnetic insulators such as yttrium-iron garnet (YIG) \cite{Cherepanov93}.
Although previously many 
authors have used S-theory and extensions thereof to study
the physics of
pumped magnon gases, \cite{Cherepanov93,Araujo74,Tsukernik75,Vinikovetskii79,Lavrinenko81,Zvyagin82,Zvyagin85,Lim88,Kalafati89,Lvov94,Zvyagin07,Rezende09,Kloss10,Safonov13,Slobodianiuk17} the available theoretical descriptions are still incomplete. 
In particular, we believe that the  
following three points deserve further theoretical attention:

1.) {\it{Magnon condensation:}}
In a recent series of experiments the phenomenon of ``Bose-Einstein condensation'' of magnons in thin films of YIG has been 
observed \cite{Demokritov06,Demidov07,Dzyapko07,Demidov08a,Demokritov08,Demidov08b,Serga14,Clausen15a,Clausen15b}.
Although this phenomenon can be explained using the classical 
stochastic Landau-Lifshitz-Gilbert equation,\cite{Rueckriegel15}
a microscopic quantum mechanical description  of the experiments  based on a kinetic equation
for the magnon distribution including proper collision integrals has not been achieved.
In such an approach the original spin model should be bosonized 
using  the Holstein-Primakoff transformation \cite{Holstein40} and the kinetic equations 
for magnon distribution derived from the effective boson model should be solved  together with the 
equation of motion of the expectation values
of the magnon annihilation and creation operators \cite{Rueckriegel12}
 \begin{equation}
 \psi_{\bd{k}} ( t ) = \langle a_{\bd{k}} ( t ) \rangle, \; \; \;
\psi^{\ast}_{\bd{k}} ( t ) = \langle a^{\dagger}_{\bd{k}} ( t ) \rangle.
 \label{eq:psidef}
\end{equation}
However, in this case  the distribution functions $n_{\bd{k}} ( t )$ and
$p_{\bd{k}} ( t )$  contain redundant information and one should consider instead 
their connected counter-parts\cite{Kloss10,Fricke96}
 \begin{subequations}
  \begin{eqnarray}
 n^c_{\bd{k}} ( t ) & = & \langle \delta a^{\dagger}_{\bd{k}} ( t )
 \delta a_{\bd{k}} ( t ) \rangle
 \nonumber
 \\
 & = &  \langle  a^{\dagger}_{\bd{k}} ( t )
  a_{\bd{k}} ( t ) \rangle -
\langle  a^{\dagger}_{\bd{k}} ( t ) \rangle
 \langle  a_{\bd{k}} ( t ) \rangle,
 \label{eq:nckdef}
 \\
p^c_{\bd{k}} ( t ) & = & \langle \delta a_{- \bd{k}} ( t ) \delta a_{\bd{k}} ( t ) 
 \rangle 
 \nonumber
 \\
& = & \langle  a_{- \bd{k}} ( t )  a_{\bd{k}} ( t )  \rangle -   
 \langle  a_{- \bd{k}} ( t )  \rangle  \langle a_{\bd{k}} ( t )  \rangle,
 \label{eq:pckdef}
 \end{eqnarray}
 \end{subequations}
where $\delta a_{\bd{k}} ( t ) = a_{\bd{k}} ( t ) - \langle a_{\bd{k}} (t) \rangle$.
In conventional  S-theory \cite{Zakharov70} the
expectation values $\langle a_{\bd{k}} ( t ) \rangle$
and  $\langle a^{\dagger}_{\bd{k}} ( t ) \rangle$ are not explicitly taken into account,
so that the kinetic equations of S-theory should be modified if the
expectation values of the
magnon operators  are finite.
In this work we show how the kinetic equations of S-theory should be modified
in the presence of magnon condensation.

2.) {\it{Mode-decoupling:}} 
If the collision integrals in the kinetic equations for the magnon distribution are
completely neglected,
S-theory reduces to
the time-dependent self-consistent Hartree-Fock approximation.
Interaction effects  are then  taken into account  
via a time-dependent magnon self-energy $\Sigma_{\bd{k}} ( t )$  which depends on the
distribution function $n_{\bd{q}} ( t )$ for all momenta $\bd{q}$,
 \begin{eqnarray}
 \Sigma_{\bd{k}} ( t ) & = & \frac{1}{N} \sum_{\bd{q}} T_{ \bd{k} , \bd{q}} n_{\bd{q}} ( t ),
 \label{eq:SigmaS}
 \end{eqnarray}
where the $T_{ \bd{k} , \bd{q}}$ is a matrix element of the
two-magnon interaction vertex (see Eq.~\eqref{eq:Tkq} below) and $N$ is the number of lattice sites.
The resulting kinetic equation is then an integro-differential equation, which will be discussed
in detail in Sect.~\ref{sec:kinetic_eq}. 
To reduce  the mathematical complexity of the problem, 
it has been proposed\cite{Zakharov70,Zakharov74,Araujo74}
to retain only the term  with $\bd{q} = \bd{k}$ in the sum 
of Eq.~(\ref{eq:SigmaS}), which 
amounts to replacing the self-consistent Hartree-Fock self-energy by
 \begin{equation}
 \Sigma_{\bd{k}} ( t ) \rightarrow  \frac{1}{N} T_{ \bd{k} , \bd{k}} n_{\bd{k}} ( t ).
 \label{eq:Sigmasub}
 \end{equation}
The authors of Refs.~\cite{Zakharov70,Zakharov74,Araujo74}]  have argued that
this decoupling of modes is justified to describe a stationary state where
only the mode with the smallest damping (which was introduced phenomenologically)
is significantly populated. 
If the damping of all other modes is sufficiently strong, their contribution to the sum
\eqref{eq:SigmaS}  can be neglected for long times, so that
only a single term with the smallest damping survives.
Although Zahkarov {\it{et al.}}\cite{Zakharov70,Zakharov74} have given intuitive arguments
how the damping selects the appropriate mode leading to Eq.~\eqref{eq:Sigmasub},  
it is somewhat  unsatisfying  that the damping was introduced phenomenologically.
If the damping indeed has the assumed form,   
the resulting stationary magnon distribution  is a reasonable approximation to 
the stationary non-equilibrium distribution.
On the other hand, the mode decoupling assumed in Eq.~(\ref{eq:Sigmasub}) 
cannot be justified to describe
 the magnon kinetics at finite times.
Indeed, we will show in Sect.~\ref{sec:dynamics} that 
the substitution (\ref{eq:Sigmasub}) does not give a
quantitatively accurate description of the time evolution of the magnon distribution.

3.) {\it{Microscopic collision integral:}} 
Although the collision integral in the kinetic equation for the magnon distribution has been written down in the Born approximation \cite{Vinikovetskii79},  and the effect of collisions
on the non-equilibrium magnon dynamics has been taken into account phenomenologically by introducing (by hand) a relaxation rate into the kinetic equations of S-theory, we have not been able to find in the literature 
an explicit solution of the quantum kinetic equation
for the magnon distribution including the microscopic
expressions for the collision integrals.
This technically extremely challenging problem is 
beyond the scope of this work~\cite{footnoteyig}. 
The collision integrals are of a more complex form than the self-energy $\Sigma_\kv(t)$ given by Eq.~\eqref{eq:SigmaS} as they involve two momentum integrations and depend on the non-equilibrium distribution functions. In the stationary non-equilibrium state the explicit calculation of the collision integrals is therefore numerically very challenging. It turns out that the microscopic collision integrals as a function of the external magnetic field strength display peaks at certain field strengths leading to enhanced or reduced magnon densities \cite{footnoteyig}.

The rest of this article is organized as follows.
In Sect.~\ref{sec:eff_H} we briefly summarize the derivation of a microscopic
boson Hamiltonian describing the pumped magnon gas in a ferromagnetic insulator such as YIG. 
In Sect.~\ref{sec:kinetic_eq} we derive the kinetic equations describing 
the time-evolution of the magnon distribution  using  different levels of approximation.
In particular, we derive the modifications of  S-theory in the presence of magnon condensation. 
The stationary non-equilibrium solution of these equations are derived in 
Sect.~\ref{sec:stat}, while in Sect.~\ref{sec:dynamics} 
the resulting time-evolution of the magnon distribution
is calculated numerically to assess  the effect of magnon condensation 
 and the
effect of the mode-decoupling approximation (\ref{eq:SigmaS})
on the time-evolution of the distribution function within S-theory.
Our main results are summarized in Sect.~\ref{sec:conclusions}. 
To make our work self-contained, we describe in the appendix the derivation of the effective magnon Hamiltonian and the two-magnon interaction vertices specifically for YIG.

\section{\label{sec:eff_H}
Hamiltonian for pumped magnons in YIG}
To set up our notation, 
let us briefly outline the derivation of the effective Hamiltonian describing the pumped magnon gas in YIG
starting from the time-dependent spin 
Hamiltonian \cite{Cherepanov93,Lvov94,Rezende09,Kloss10,Hick10,Rueckriegel14,Rezende06,Kreisel09}
\begin{eqnarray}
{\mathcal{H}}(t)&=&-\frac{1}{2}\sum\limits_{i j}\sum\limits_{\alpha \beta}\left[J_{i j}\delta^{\alpha \beta} + D_{i j}^{\alpha \beta}\right]S_i^\alpha S_j^\beta \nonumber\\*
 &&-\left[h_0+h_1\cos\left(\omega_0 t\right)\right]\sum\limits_i S_i^z,
\label{eq:eff_H}
\end{eqnarray}
where $i, j = 1, \dots, N$ label the sites $\bd{r}_i$ and $\bd{r}_j$ of a cubic lattice with $N$ sites  and lattice spacing $a \approx 12.376$\AA,
and $\alpha, \beta = x, y, z$ denote to the three components of the spin operators $S_i^\alpha$.
The exchange coupling $J_{ij}=J(\bd{r}_i-\bd{r}_j)$ 
between nearest neighbors has the numerical value $J\approx1.29$K.
The energy scales $h_0 = \mu H_0$ and $h_1 = \mu H_1$ (where $\mu = g \mu_B$) 
represent the Zeeman energies associated with a static magnetic field $ H_0 $ 
and a time dependent field $H_1$ oscillating with frequency $\omega_0$.
As explained in  Refs.~\cite{Kreisel09,Tupitsyn08}, we may set the $g$-factor equal to two and work with an effective spin $S\approx14.2$. 
We restrict ourselves to the description of an infinitely long stripe aligned with the $z$-axis 
of width $w$ and thickness $d = N a$. For experimentally relevant YIG stripes this thickness is several thousand lattice spacings. The dipolar tensor $D_{ij}^{\alpha\beta}=D^{\alpha\beta}(\bd{r}_i-\bd{r}_j)$ can be written as \cite{Kreisel09,Filho00}
\begin{equation}
 D_{i j}^{\alpha \beta} = \left(1-\delta_{i j}\right)\frac{\mu^2}{\left|\bd{r}_{i j}\right|^3}\left[3\hat{{r}}_{i j}^\alpha\hat{{r}}_{i j}^\beta-\delta^{\alpha\beta}\right] ,
\end{equation}
where $\bd{r}_{ij}=\bd{r}_i-\bd{r}_j$ and $\hat{\bd{r}}_{ij}=\bd{r}_{ij}/\left|\bd{r}_{ij}\right|$.  
We assume that the classical ground state is a saturated ferromagnet for magnetic fields oriented along the direction of the stripe.
In the ground state all spins align with the static magnetic field which defines the direction of the macroscopic magnetization.
The spin Hamiltonian (\ref{eq:eff_H}) can then be bosonized via the Holstein-Primakoff transformation \cite{Holstein40} and the resulting effective boson Hamiltonian can be expanded in powers of the small parameter $1/S$,
\begin{equation}
 \mathcal{H}(t)=\mathcal{H}_0(t)+\mathcal{H}_2(t)+\mathcal{H}_3+\mathcal{H}_4+\mathcal{O}(S^{-1/2}),
 \label{eq:H_1/S}
\end{equation}
where the $\mathcal{H}_n$ contain all terms of order $n$ in the boson operators.
We apply a partial Fourier transformation in the $yz$-plane assuming the width $w$ of the sample is infinite, which is reasonable because for experimentally relevant geometries the width $w$ is much larger than the thickness $d$.
As explained in the appendix (see also Refs.~\cite{Kloss10,Hick10,Kreisel09})
the time-independent  part of the off-diagonal terms in the quadratic part $\mathcal{H}_2 (t)$
of the effective boson Hamiltonian are then eliminated by means of a Bogoliubov 
transformation and the non-resonant terms in the remaining time-dependent part of
 $\mathcal{H}_2 (t)$ are simply dropped. With these approximations
\begin{eqnarray}
 \mathcal{H}_2\left(t\right) &=& \sum\limits_\kv \left[\varepsilon_\kv a^\dagger_\kv a_\kv 
 + \frac{1}{2} V_\kv \text{e}^{-i \omega_0 t} a^\dagger_\kv a^\dagger_{-\kv}
\right.
\nonumber\\*
 &&
\hspace{18mm} + \left.  \frac{1}{2} V_\kv^* \text{e}^{i \omega_0 t} a_{-\kv} a_\kv\right], 
\label{eq:H_2}
\end{eqnarray}
where the boson operator $a^{\dagger}_{\bd{k}}$ creates a magnon with energy
$\varepsilon_\kv$. In the long wavelength limit $\varepsilon_{\bd{k}}$ 
 can be approximated by \cite{Kreisel09,Tupitsyn08,Kalinkos86}
\begin{equation}
 \varepsilon_\kv=\sqrt{\left[h_0+\rho \kv^2+\left(1-f_\kv\right)\Delta\sin^2\theta_\kv\right]\left[h_0+\rho\kv^2+f_\kv\Delta\right]},
 \label{eq:disp_rel}
\end{equation}
where the spin stiffness associated with the exchange energy is denoted by
 $\rho=JSa^2$,
and the energy scale associated with the dipolar energy is
$\Delta= 4\pi\mu^2 S / a^3$.
The in-plane wavevector $\bd{k}$ is parametrized as
$
\kv = k_z \bd{e}_z + k_y \bd{e}_y = |\kv| \left(\cos\theta_\kv \bd{e}_z + \sin\theta_\kv \bd{e}_y\right)$,
so that
$\theta_{\bd{k}}$ is the angle between the in-plane wavevector $\bd{k}$  and 
the external magnetic field.
The form factor $f_\kv$ can be approximated by \cite{Kreisel09}
\begin{equation}
f_\kv = \frac{1 - \text{e}^{-\left|\kv\right|d}}{\left|\kv\right|d},
 \label{eq:formdef}
\end{equation}
and the  pumping energy $V_\kv$ is \cite{Kloss10,Hick10,Rueckriegel14}
\begin{equation}
 V_\kv=\frac{h_1\Delta}{4\varepsilon_\kv}\left[-f_\kv+\left(1-f_\kv\right)\sin^2\theta_\kv\right].
 \label{eq:pump_en}
\end{equation}
We can remove the explicit time-dependence of ${\cal{H}}_2 ( t )$ in Eq.~(\ref{eq:H_2})
by transforming to the rotating reference frame via the canonical transformation
\begin{eqnarray}
 \tilde{a}_\kv &=& \text{e}^{i \frac{\omega_0}{2} t} a_\kv,   \; \; \; 
 \tilde{a}^\dagger_\kv = \text{e}^{-i \frac{\omega_0}{2} t} a^\dagger_\kv.
 \label{eq:rotref}
\end{eqnarray}
The quadratic part of the transformed Hamiltonian in the
rotating reference frame is then independent of time,
\begin{equation}
 \tilde{\mathcal{H}}_2 = \sum\limits_\kv \left[E_\kv \tilde{a}_\kv^\dagger \tilde{a}_\kv + \frac{V_\kv}{2} \tilde{a}_\kv^\dagger \tilde{a}_{-\kv}^\dagger  + \frac{V_\kv^*}{2} \tilde{a}_{-\kv} \tilde{a}_\kv\right],
 \label{eq:H2_t}
\end{equation}
\sloppy
with shifted magnon energy $E_\kv = \varepsilon_\kv - \omega_0/2$.
Unfortunately,  the canonical transformation (\ref{eq:rotref})
generates some explicit time-dependence in the
 interaction part of the Hamiltonian \cite{Hick10}. For example,
the cubic part $\tilde{\cal{H}}_3 ( t ) $ of the magnon Hamiltonian contains terms of the form
$e^{ i \omega_0 t /2 } \tilde{a}^{\dagger}_{- \bd{k}_1} 
\tilde{a}^{\dagger}_{- \bd{k}_2} \tilde{a}_{ \bd{k}_1 + \bd{k}_2 }$.
For our purpose we can simply neglect $\tilde{\cal{H}}_3 ( t )$ because
it involves only rapidly oscillating terms with frequency $\omega_0/2 \gg |\epsilon_\kv-\omega_0/2|$ which should be consistently dropped in rotating-wave approximation.
Finally, the quartic part of the mag\-non Hamiltonian in the
rotating reference frame is of the form~\cite{Hick10}
\begin{eqnarray}
 \tilde{\mathcal{H}}_4 (t ) &=& \frac{1}{N} \sum\limits_{\kv_1, \dots, \kv_4} \delta_{\kv_1+\dots+\kv_4, 0} \left[\frac{1}{\left(2!\right)^2} \Gamma^{\bar{a} \bar{a} a a}_{1, 2; 3, 4} 
\tilde{a}^\dagger_{-1} \tilde{a}^\dagger_{-2} \tilde{a}_3 \tilde{a}_4 \right.\nonumber\\*
&&\left. + \frac{1}{3!} e^{ - i \omega_0 t} \Gamma^{\bar{a} a a a}_{1; 2, 3, 4} 
 \tilde{a}^\dagger_{-1} \tilde{a}_2 \tilde{a}_3 \tilde{a}_4 
 \right.\nonumber
 \\*
 && \left. + \frac{1}{3!} e^{ i \omega_0 t} 
 \Gamma^{\bar{a} \bar{a} \bar{a}a}_{1, 2, 3; 4} \tilde{a}^\dagger_{-1} \tilde{a}^\dagger_{-2} 
 \tilde{a}^\dagger_{-3} \tilde{a}_4 \right.\nonumber\\*
 &&\left. + \frac{1}{4!} e^{ - 2 i \omega_0 t}
\Gamma^{aaaa}_{1, 2, 3, 4} \tilde{a}_1 \tilde{a}_2 \tilde{a}_3 \tilde{a}_4 \right.\nonumber\\*
 &&\left. + \frac{1}{4!} e^{ 2 i \omega_0 t} \Gamma^{\bar{a} \bar{a} \bar{a}\bar{a}}_{1, 2, 3, 4} 
\tilde{a}^\dagger_{-1} \tilde{a}^\dagger_{-2} \tilde{a}^\dagger_{-3} \tilde{a}^\dagger_{-4} \right] ,
 \label{eq:H4rot}
\end{eqnarray}
\fussy
where the Kronecker-$\delta$ enforces momentum conservation 
and we use the abbreviation $\tilde{a}_1 = \tilde{a}_{\bd{k}_1 }$.
Explicit expressions for the
two-magnon interaction vertices appearing in Eq.~(\ref{eq:H4rot})
are given in the appendix, see Eqs.~\eqref{eq:vertices1}--\eqref{eq:vertices5}.
Within the rotating-wave approximation, it is consistent to drop
all  oscillating terms on the right-hand side of Eq.~(\ref{eq:H4rot}).

\section{\label{sec:kinetic_eq}
Collisionless kinetic equations}

In this section we derive kinetic equations for the magnon distribution in the rotating reference frame using three different levels of approximation. 
However, a proper microscopic treatment of the
collision integrals is beyond the scope of this work \cite{footnoteyig}.
Note that the phase factor $e^{ \pm i \omega_0 t /2}$
in Eq.~(\ref{eq:rotref}) cancels in the diagonal distribution $n_{\bd{k}} ( t )$, so that
in the rotating reference frame
the diagonal and the off-diagonal distribution functions are given by
 \begin{subequations}
\begin{eqnarray}
 n_\kv(t) &=&     \langle \tilde{a}^\dagger_\kv(t) \tilde{a}_\kv(t)\rangle = \langle a^\dagger_\kv(t) a_\kv(t)\rangle, 
 \\
\tilde{p}_\kv(t) &=& \langle \tilde{a}_{-\kv}(t) \tilde{a}_\kv(t)\rangle = \mbox{e}^{i \omega_0 t} p_\kv(t).
\end{eqnarray}
 \end{subequations}

\subsection{Non-interacting system}
To begin with, let us completely neglect all magnon-mag\-non interactions 
and approximate the Hamiltonian
for the pumped magnon gas in the rotating reference frame by the quadratic 
Hamiltonian
 $\tilde{\mathcal{H}}_2$ in Eq.~\eqref{eq:H2_t}. 
From the  Heisenberg equations of motion in the rotating reference frame,
 \begin{equation}
 i \partial_t \tilde{a}_{\bd{k}} = \bigl[ \tilde{a}_{\bd{k}} , \tilde{\cal{H}} ( t ) \bigr],
 \; \; \; 
 i \partial_t \tilde{a}^{\dagger}_{\bd{k}} = \bigl[ \tilde{a}^{\dagger}_{\bd{k}} , \tilde{\cal{H}} ( t ) \bigr],
 \end{equation}
it is then easy to show that
\begin{eqnarray}
 \partial_t n_\kv(t) + i \left[V_\kv \tilde{p}^*_\kv(t) - V^*_\kv \tilde{p}_\kv(t)\right] &=& 0, \label{eq:n_raw}\\
 \partial_t \tilde{p}_\kv(t) + 2 i E_\kv \tilde{p}_\kv(t) + i V_\kv \left[2 n_\kv(t) + 1\right] &=& 0, \label{eq:p_raw}
\end{eqnarray}
where in the second equation we have used  $E_\kv = E_{-\kv}$ and $n_\kv = n_{-\kv}$.
These equations can be solved exactly \cite{Kloss10}.
In the regime    $|E_\kv| > |V_\kv|$ the solutions exhibit an oscillatory behavior, while in the
strong-pumping regime where $|V_\kv| > |E_\kv|$ the solutions grow 
exponentially in time \cite{Suhl57,Schloemann60,Kloss10}.

\subsection{\label{sec:S-theory} S-theory: 
time-dependent self-consistent Hartree-Fock approximation}

Interactions between magnons eventually lead to a saturation of the 
exponential growth of the magnon distribution in the strong-pumping regime.
The simplest approximation which includes this interaction-induced saturation mechanism is
the time-dependent self-consistent Hartree-Fock approximation, which in this context is 
called S-theory \cite{Zakharov70,Zakharov74,Lim88,Lvov94,Rezende09}.
 The kinetic equations \eqref{eq:n_raw} and \eqref{eq:p_raw}  are then replaced by
\begin{eqnarray}
 \partial_t n_\kv(t) + i \left[\tilde{V}_\kv (t) \tilde{p}^*_\kv(t) - \tilde{V}^*_\kv (t) \tilde{p}_\kv(t)\right] &=& 0, \hspace{7mm} \label{eq:n_S} \\
 \partial_t \tilde{p}_\kv(t) + 2 i \tilde{E}_\kv ( t ) \tilde{p}_\kv(t) + i \tilde{V}_\kv (t) \left[2 n_\kv(t) + 1\right] &=& 0, \label{eq:p_S}
\end{eqnarray}
where the renormalized magnon dispersion $\tilde{E}_\kv(t)$ and pumping energy $\tilde{V}_\kv(t)$ depend now on the distribution functions,
\begin{eqnarray}
 &&\tilde{E}_\kv(t) = E_\kv + \frac{1}{N} \sum\limits_\qv\Bigl[\Gamma^{\bar{a}\bar{a} aa}_{-\kv, -\qv; \qv, \kv} n_\qv(t)  \nonumber\\*
 && + \frac{1}{2} \text{e}^{-i \omega_0 t} \Gamma^{\bar{a}aaa}_{-\kv; -\qv, \qv, \kv} \tilde{p}_\qv(t) + \frac{1}{2} \text{e}^{i \omega_0 t} \Gamma^{\bar{a} \bar{a} \bar{a}a}_{-\kv, -\qv, \qv; \kv} \tilde{p}^*_\qv(t)\Bigr], \nonumber\\\label{eq:E_renorm}\\
 &&\tilde{V}_\kv(t) = V_\kv + \frac{1}{2 N} \sum\limits_\qv\Bigl[
 \text{e}^{i \omega_0 t} \Gamma^{\bar{a}\bar{a}\bar{a} a}_{-\kv, \kv, -\qv; \qv} n_\qv(t) 
 \nonumber\\*
 &&+ \frac{1}{2} \Gamma^{\bar{a} \bar{a} aa}_{-\kv, \kv; -\qv, \qv} \tilde{p}_\qv(t) + \frac{1}{2} \text{e}^{2 i \omega_0 t} \Gamma^{\bar{a}\bar{a}\bar{a}\bar{a}}_{-\kv, \kv, -\qv; \qv} \tilde{p}^*_\qv(t)\Bigr]. \label{eq:V_renorm}
\end{eqnarray}
Within the rotating-wave approximation we can drop the rapidly 
oscillating terms proportional to $\text{e}^{\pm i \omega_0 t}$ and $\text{e}^{\pm 2 i \omega_0 t}$, so that the above self-consistency equation reduce to
 \begin{subequations}
\begin{eqnarray}
 \tilde{E}_\kv(t) &=& E_\kv + \frac{1}{N} \sum\limits_\qv T_{\kv,\qv} n_\qv(t), \label{eq:E_r_ex}\\
 \tilde{V}_\kv(t) &=& V_\kv + \frac{1}{2 N} \sum\limits_\qv S_{\kv,\qv} \tilde{p}_\qv(t), \label{eq:V_r_ex}
\end{eqnarray}
 \label{eq:HFenergies}
 \end{subequations}
with
\begin{subequations}
\begin{eqnarray}
 T_{\kv,\qv} &=& \Gamma^{\bar{a} \bar{a}aa}_{-\kv,-\qv;\qv,\kv}, \label{eq:Tkq}\\
 S_{\kv,\qv} &=& \Gamma^{\bar{a}\bar{a}aa}_{-\kv,\kv;-\qv,\qv}.
\end{eqnarray}
\end{subequations}

\subsection{\label{sec:mode_decoupling}
Mode decoupling}
\sloppy
At this point the kinetic equations are still non-trivial integro-differential equations. However, the authors of Refs.~\cite{Zakharov70,Zakharov74,Araujo74}
have  proposed to decouple the modes with different wavevectors 
by approximating the sums in Eqs.~\eqref{eq:E_r_ex} and \eqref{eq:V_r_ex} by the single term 
where the loop momentum $\qv$ is equal to the external momentum $\kv$, 
\begin{subequations}
\begin{eqnarray}
 \tilde{E}_\kv(t) &\approx& E_\kv + \frac{1}{N} T_{\kv, \kv} n_\kv(t), \label{eq:sum_approx-a}\\
 \tilde{V}_\kv(t) &\approx& V_\kv + \frac{1}{2 N} S_{\kv, \kv} \tilde{p}_\kv(t). \label{eq:sum_approx-b}
\end{eqnarray}
\label{eq:sum_approx}
\end{subequations}
\fussy
As already mentioned in the introduction after Eq.~(\ref{eq:SigmaS}), this substitution has been 
justified by observing that after sufficiently long times only the mode with the longest lifetime is significantly populated, so that
the momentum dependence of the magnon damping is crucial to justify Eq.~(\ref{eq:sum_approx}).
This substitution can therefore only be used to describe the asymptotic 
long-time dynamics of the magnon distribution, including  a possible stationary non-equilibri\-um state.

\subsection{\label{sec:Smag}
S-theory with magnon condensation}
\fussy
In the previous subsection, we have not taken into account the possibility that the non-equilibrium
expectation values $\psi_{\bd{k}} ( t ) = \langle a_{\bd{k}} ( t ) \rangle$ of the magnon operators are finite, see 
Eq.~(\ref{eq:psidef}).
In the strong-pumping regime or in the regime where the magnons condense, the finite value of 
$\psi_{\bd{k}} ( t )$
cannot be ignored, so that we should complement our system of kinetic equations by an equation of motion
for $\psi_{{\bd{k}}} ( t )$, which is the analogue of  the Gross-Pitaevskii equation describing
the dynamics of the order parameter of a superfluid. 
The parametrization of the dynamics in terms of the 
correlation functions $n_{\bd{k}} ( t )$ and $\tilde{p}_{\bd{k}} (t)$ is then redundant 
and it is better to consider their connected counter-parts in the rotating reference frame,\cite{Kloss10}
 \begin{subequations}
\begin{eqnarray}
 n^c_\kv(t) = \langle\delta \tilde{a}^\dagger_\kv(t) \delta \tilde{a}_\kv(t)\rangle, \\
 \tilde{p}^c_\kv(t) = \langle\delta \tilde{a}_{-\kv}(t) \delta \tilde{a}_\kv(t)\rangle,
\end{eqnarray}
 \end{subequations}
where $\delta \tilde{a}_\kv(t) = \tilde{a}_\kv(t) - \langle \tilde{a}_\kv(t) \rangle$.
Note that conventional S-theory does not explicitly take finite expectation values of the magnon 
annihilation and creation operators into account. Defining the expectation value of the magnon operators in the
rotating reference frame,
\begin{equation}
 \tilde{\psi}_{\kv}(t) = \langle\tilde{a}_\kv(t)\rangle ,
\end{equation}
\sloppy
we obtain within the  self-consistent time-dependent Hartree-Fock approximation and the 
rotating-wave approximation,
\begin{eqnarray}
 \partial_t n_\kv^c + i \left[\tilde{V}_\kv   \left(\pvf{\kv}\right)^* - \tilde{V}_\kv^*  \pvf{\kv}\right] &=& 0, \label{eq:dgl_n0} \\
 \partial_t \pvf{\kv} + 2 i \tilde{E}_\kv   \pvf{\kv} + i \tilde{V}_\kv   \left[2 n_\kv^c + 1\right] &=& 0, \label{eq:dgl_p0} \\
 \partial_t \tilde{\psi}_{\kv} + i \tilde{E}_\kv \tilde{\psi}_{\kv}  + i \tilde{V}_\kv 
\tilde{\psi}^{\ast}_{-\kv} &=& 0, \label{eq:dgl_a0}
\end{eqnarray}
where it is understood that all quantities are time-dependent and the renormalized magnon dispersion and pumping energy are now given by
\begin{subequations}
\begin{eqnarray}
 \tilde{E}_\kv &=& E_\kv + \frac{1}{N} \sum\limits_\qv T_{\kv, \qv} 
\Bigl( n^c_\qv + 
 \bigl| \tilde{\psi}_{\qv} \bigr|^2 \Bigr),\\
 \tilde{V}_\kv   &=& V_\kv + \frac{1}{2 N} \sum\limits_\qv S_{\kv, \qv} \left(\tilde{p}_\qv^c + 
\tilde{\psi}_{-\qv}\tilde{\psi}_{\qv}\right).
\end{eqnarray}
\label{eq:EVint}
\end{subequations}

\section{\label{sec:stat}
Stationary non-equilibrium state}
\fussy
Within the approximations described in 
Sect.~\ref{sec:S-theory} we can find a stationary non-equilibrium solution 
for the distribution functions. We further simplify the self-consistent energy $\tilde{E}_\kv(t)$ and the pumping $\tilde{V}_\kv(t)$ using the mode-decoupling approximation
described by Eqs.~\eqref{eq:sum_approx-a} and \eqref{eq:sum_approx-b}. 
We also approximate $2 n^c_\kv+1\approx 2n^c_\kv$ in Eq.~\eqref{eq:dgl_p0} because the magnon density in the stationary state is of order $N \gg 1$.
In the regime of parametric instability,
\begin{equation}
 \left|V_\kv\right| > \left|E_\kv\right|,
 \label{eq:ST_cond}
\end{equation}
we find the stationary solution within S-theory~\cite{Zakharov70,Zakharov74},
\begin{subequations}
\begin{eqnarray}
 n_\kv &=& N \frac{\left|V_\kv\right| - \left|E_\kv\right|}{T_{\kv, \kv} + \frac{1}{2} S_{\kv, \kv}},
 \label{eq:n_stat_S} \\
 \tilde{p}_\kv &=&  - n_\kv.
\end{eqnarray}
\label{eq:stat_S}
\end{subequations}
Eq.~\eqref{eq:ST_cond} implies that there is a minimum strength of the pumping field $h_1$ which 
is necessary to macroscopically populate magnons with wavevector $\kv$ and energy $\varepsilon_\kv$,
\begin{equation}
 h_{1, \text{min}}(\kv) = \frac{4}{\Delta} \left|\frac{\varepsilon_\kv\left(\varepsilon_\kv - \frac{\omega_0}{2}\right)}{-f_\kv+\left(1-f_\kv\right)\sin^2\theta_\kv}\right|.
 \label{eq:h1_min}
\end{equation}
Note that at this point we have neglected magnon damping 
which would lead to a relaxation of magnon occupation.
Without damping, this equation suggests that mag\-nons satisfying $\epsilon_\kv=\omega_0/2$ are always excited for infinitesimal small pumping fields $H_1$.

The above results for the stationary magnon distributions do not change if 
we use the modified S-theory including magnon condensation 
as described in Sect.~\ref{sec:Smag}.
In this case we obtain from \eqref{eq:dgl_n0}--\eqref{eq:dgl_a0} with 
$\partial_t n^c_\kv = \partial_t \tilde{p}^c_\kv = \partial_t \vfv{\kv} = 0$,
\begin{subequations}
\begin{eqnarray}
 n_{\kv} =   n^{c}_\kv+ \bigl|\tilde{\psi}_{\bd{k}} \bigr|^2 &=& N \frac{\left|V_\kv\right| - \left|E_\kv\right|}{T_{\kv, \kv} + \frac{1}{2} S_{\kv, \kv}}, \label{eq:n_stat}\\
 \tilde{p}_{\kv} = \tilde{p}^{c}_\kv+\bigl( \tilde{\psi}_{\kv} \bigr)^2 &=& -n^{c}_\kv
 -\bigl| \tilde{\psi}_{\kv} \bigr|^2 \label{eq:p_stat}.
\end{eqnarray}
\end{subequations}
Note that the partition of the magnon distribution function $n_\kv$ into a 
connected part $n^{c}_\kv$ and a contribution $|\vfv{\kv}|^2$ from the finite expectation value of the magnon operators
is ambiguous as long as the magnon damping is neglected.

Another way to derive the stationary magnon distribution functions
in Eqs.~\eqref{eq:stat_S} is to assume that in the stationary state only
one pair of  magnons with momenta $\pm\kv$ is significantly occupied. 
This assumption was justified 
in Refs.~\cite{Zakharov70,Zakharov74} by invoking again the
phenomenologically introduced magnon damping, which
leads to a decoupling of magnon modes. 
If we initially prepare the system in a state where more than one pair of $\kv$ modes is significantly occupied,
the usual argument is that after sufficiently long times only the pair of modes
with the smallest damping will survive.
A simple phenomenological way to model the effect of magnon damping 
is by inserting (by hand) 
a damping rate $\gamma_\kv$ into the equations of motion 
of the magnon creation and annihilation operators.  In the rotating reference frame
the resulting modified equations of motion are
\begin{subequations}
 \begin{eqnarray}
  \partial_t\tilde{a}_\kv(t)&=& \left(- iE_\kv - \gamma_\kv\right)\tilde{a}_\kv-iV_\kv\tilde{a}_{-\kv}^\dagger, \\
  \partial_t\tilde{a}_\kv^\dagger(t)&=&\left( iE_\kv -\gamma_\kv\right) \tilde{a}_\kv^\dagger
 +iV_\kv^\ast \tilde{a}_{-\kv}.
 \end{eqnarray}
\end{subequations}
For simplicity let us focus on the magnons satisfying the resonance condition
 $\varepsilon_\kv=\omega_0/2$ where  $E_\kv=0$. Then S-theory gives a 
stationary solution \cite{Zakharov70,Zakharov74}
\begin{subequations}
\begin{eqnarray}
 n_\kv &=& N \frac{\sqrt{V_\kv^2-\gamma_\kv^2}}{T_{\kv, \kv} + \frac{1}{2} S_{\kv, \kv}},
 \label{eq:n_stat_S_damp} \\
 \tilde{p}_\kv &=& - n_\kv,
\end{eqnarray}
\end{subequations}
provided the pumping compensates the losses due to damping,
\begin{equation}
 \left|V_\kv\right|>\left|\gamma_\kv\right|.
 \label{eq:ST_damp_cond}
\end{equation}
In this case a non-zero minimum strength of the pumping field $h_{1, \text{min}}({\bd k})>0$ 
is necessary to obtain a stationary state.

\begin{figure}[hptb]
 \centering
 \includegraphics[clip=true,trim=20pt 175pt 40pt 205pt,width=0.9\linewidth]{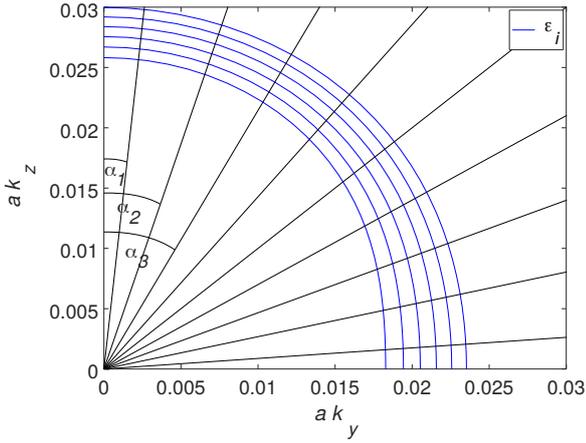}
 \caption{The discretization of momenta used for the numerical solution of the kinetic equations. $N_\epsilon=5$ energy shells with $E_{\bd{k}} \in [ - 0.1 \omega_0 , 0.1 \omega_0 ]$ are shown in the momentum space. They are further subdivided by $N_\alpha=10$ discrete angles in the interval $[ 0, \pi /2]$. The wavevectors in the center of the meshes are being used for the further calculations.}
 \label{fig:discretization}
\end{figure}

To visualize the magnon distribution in momentum space
we discretize the momenta as follows:
First of all, we retain only low-energy momenta
where $E_{\bd{k}} = \varepsilon_{\bd{k}} - \omega_0/2 \in [ - 0.1 \omega_0 , 0.1 \omega_0 ]$.
This interval is then subdivided into  $N_{\varepsilon}$ shells
$[ \varepsilon_i , \varepsilon_{ i+1} ]$ and the 
wavevectors $\bd{k}$ with $E_{\bd{k}}  \in [ \varepsilon_i , \varepsilon_{ i+1} ]$
are parametrized by $N_{\alpha}$ discrete angles   $\alpha_j$ in the interval $[ 0 , \pi /2]$
as illustrated in  Fig.~\ref{fig:discretization}.

\begin{figure}[hptb]
 \centering
 \begin{minipage}{10pt}
  (a)
 \end{minipage}
 \begin{minipage}{0.92\linewidth}
  \includegraphics[clip=true,trim=20pt 185pt 0pt 210pt,width=\linewidth]{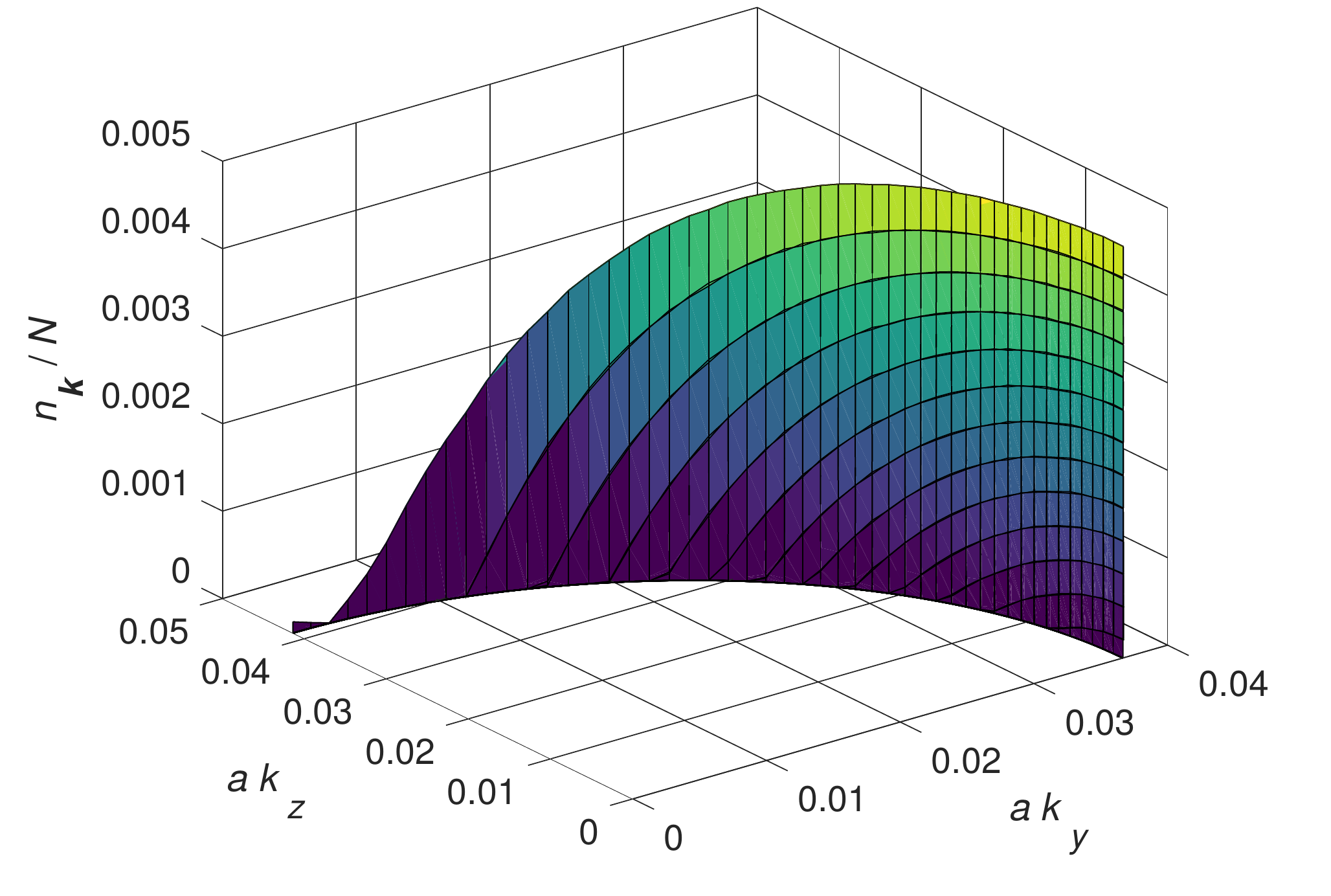}
 \end{minipage}
 \begin{minipage}{10pt}
  (b)
 \end{minipage}
 \begin{minipage}{0.92\linewidth}
  \includegraphics[clip=true,trim=20pt 175pt 65pt 205pt,width=0.9\linewidth]{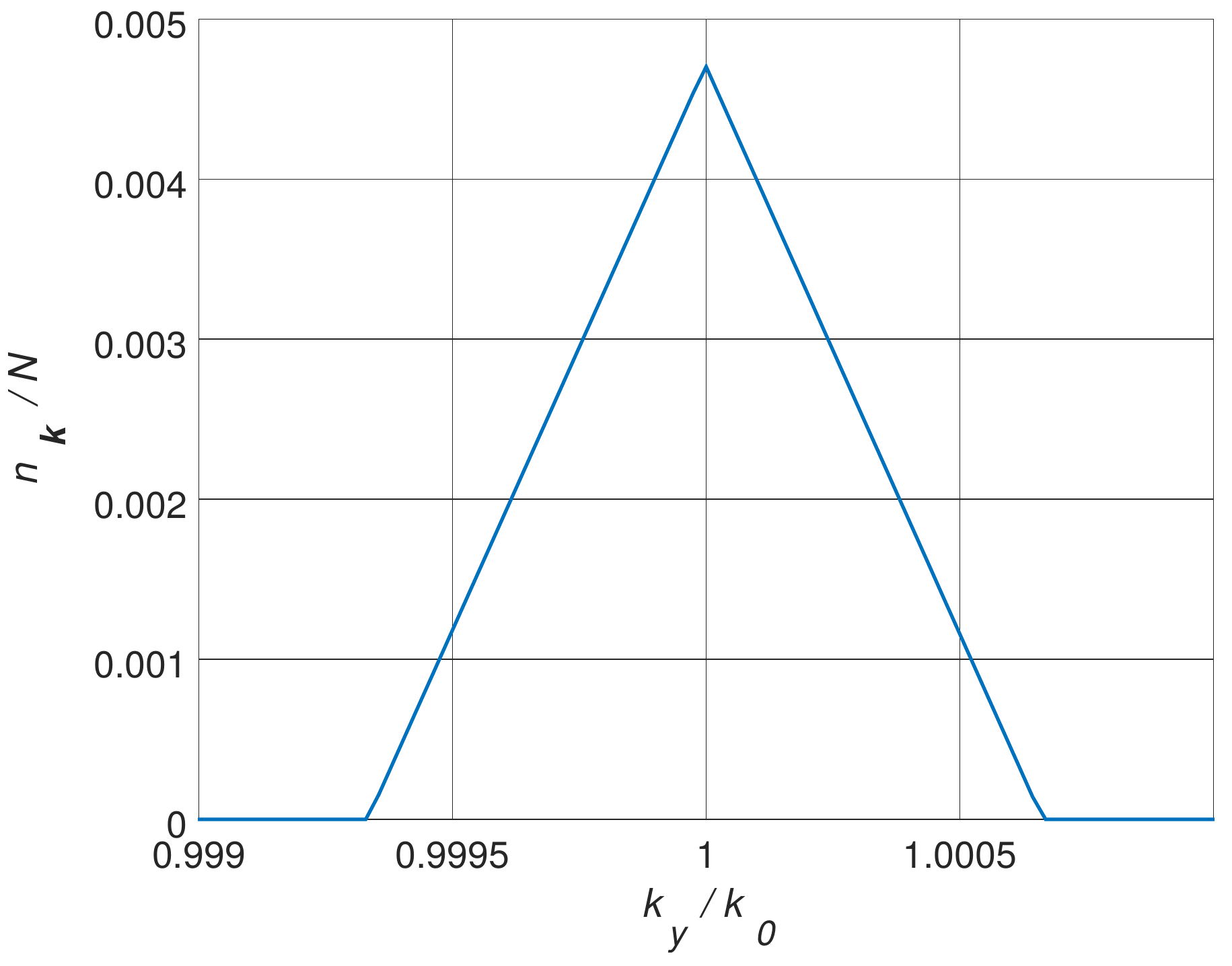}
 \end{minipage}
 \caption{(a) Plot of the magnon distribution in the stationary solution \eqref{eq:n_stat_S} within S-theory 
for a thin film of YIG with thickness $d=10\mu m$ corresponding to $N=8080$ and $H_0 = 800$ Oe, $H_1 = 40$ Oe, and $\omega_0 = 25$ GHz as a function of both components $k_y$ and $k_z$ of the in-plane momentum $\kv$, (b) as a function of $k_y$ for $k_z=0$, where $k_0=|\kv_0|$ is defined by the resonance condition $\varepsilon_{\kv_0}=\omega_0 /2$. The momenta are discretized as shown in Fig.~\ref{fig:discretization} but for different parameters.
Note that the magnon density is finite only in a small area around the resonance surface 
where the condition \eqref{eq:ST_cond} is fulfilled.}
 \label{fig:md_S_spec}
\end{figure}

\begin{figure}
 \centering
 \includegraphics[clip=true,trim=20pt 185pt 0pt 210pt,width=\linewidth]{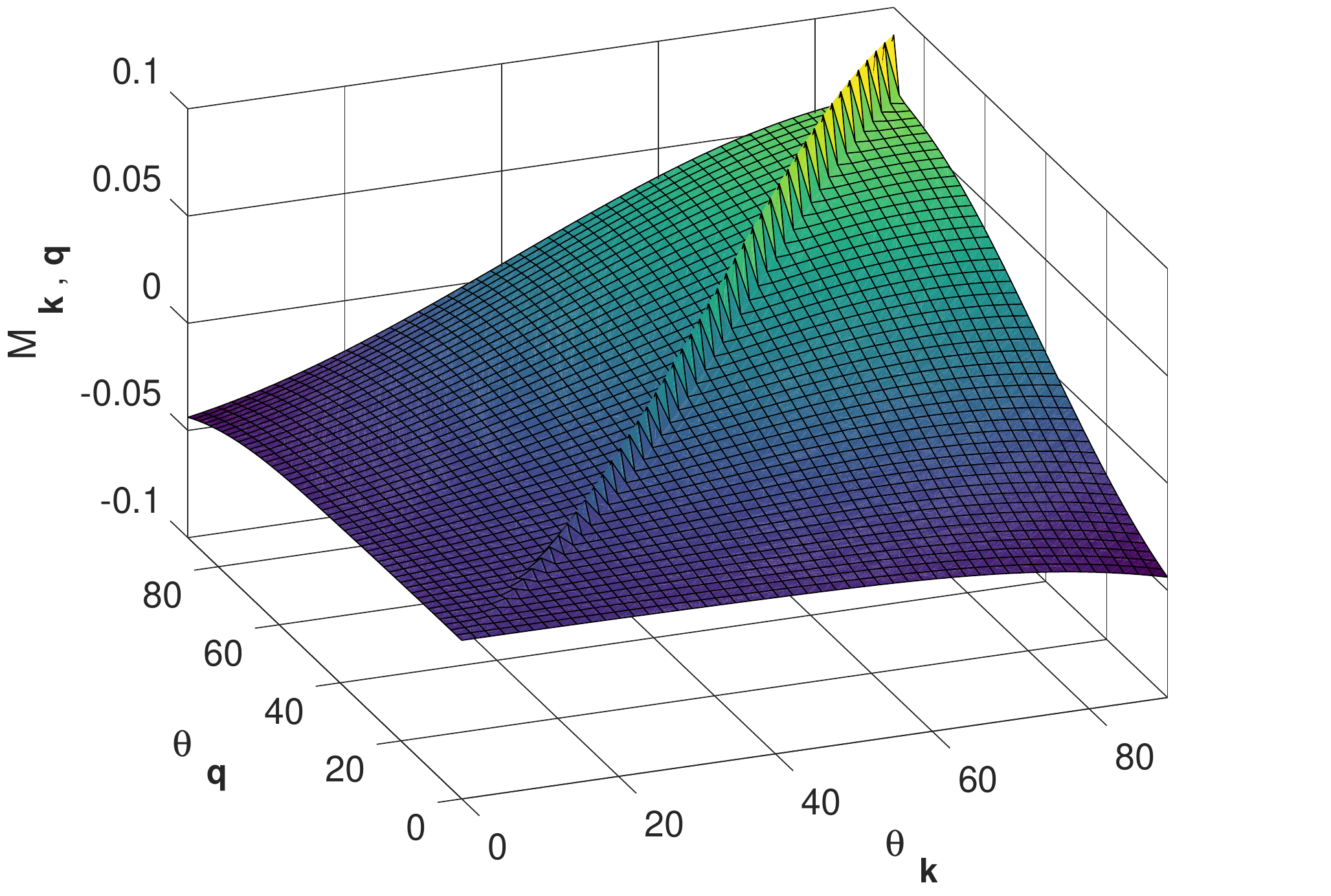}
 \caption{The dependence of $M_{\kv,\qv}=T_{\kv,\qv}+S_{\kv,\qv}/2$ on the angles $\theta_\kv$ and $\theta_\qv$ of the two wave vectors $\kv$ and $\qv$, where $\kv$ and $\qv$ both fulfill the resonance condition $\epsilon_\kv=\epsilon_\qv=\omega_0/2$.}
 \label{fig:Mkq}
\end{figure}

From the plot of the magnon distribution function obtained within S-theory in
Fig.~\ref{fig:md_S_spec} (a) we see that
for small strength 
of the pumping field $h_1 \ll h_0$ the instability condition \eqref{eq:ST_cond} 
is fulfilled only in a small region of the momentum space near the resonance surface defined by $\varepsilon_\kv=\omega_0/2$. This is also shown in Fig.~\ref{fig:md_S_spec} (b) where the magnon distribution function is plotted over $k_y$ for $k_z=0$. The shape of the distribution function is mostly determined by the numerator of Eq.~\eqref{eq:n_stat}, where $V_\kv$ is almost constant along the $y$-axis and the magnon energy depends approximately quadratically on the wavevector. The width of the distribution function is three orders of magnitude smaller than the momentum $k_0=|\kv_0|$ on the resonance surface.
This allows us to neglect the non-vanishing magnon distribution function for momenta $\kv$ which do not satisfy the resonance condition $\epsilon_\kv=\omega_0/2$.
It is therefore sufficient to take into account only momenta $\kv$ with $\varepsilon_\kv=\omega_0/2$ for calculating the stationary non-equilibrium state.
%
To quantitatively examine the accuracy of the mode-decoupling substitution 
(\ref{eq:sum_approx}) without damping,
let us try to construct a stationary solution 
of  Eq.~\eqref{eq:dgl_p0} assuming
$\tilde{p}_\kv=-n_\kv$.  In this 
case  Eq.~\eqref{eq:dgl_p0} reduces to the linear integral equation
\begin{equation}
 \frac{1}{N}
 \sum\limits_\qv  \left( T_{\kv,\qv}+\frac{1}{2}S_{\kv,\qv} \right)n_\qv 
 =\left|V_\kv\right|-E_\kv.
 \label{eq:stat_p-n}
\end{equation}
\sloppy
Obviously, a unique solution exists only 
if the matrix $M_{\kv,\qv}=T_{\kv,\qv}+\frac{1}{2}S_{\kv,\qv}$ 
is invertible. 
Numerically, we find that in certain parameter regimes this is not the case.
Moreover, even in a regime where this matrix is invertible, 
it can happen that the resulting stationary distribution is negative, which is obviously unphysical.
These problems do not arise if we retain only the diagonal elements
of the matrix $M_{\bd{k} , \bd{q} }$ in which case we recover
Eq.~(\ref{eq:n_stat_S}).
Fig.~\ref{fig:Mkq} shows $M_{\kv,\qv}$ plotted over the angles $\theta_\kv$ and $\theta_\qv$ of two vectors $\kv$ and $\qv$ with $\epsilon_\kv=\epsilon_\qv=\omega_0/2$. A peak for $\kv=\qv$ can be observed but this finding is not sufficient to justify the mode decoupling.
However, as the stationary solution obtained from the integral equation (\ref{eq:stat_p-n}) can have unphysical features,
we conclude that the kinetic equations derived in Sect.~\ref{sec:kinetic_eq} are not suitable to describe experimentally relevant systems and the replacement of the Hartree-Fock self-energies
in Eq.~(\ref{eq:HFenergies}) by the decoupled expressions
in Eq.~(\ref{eq:sum_approx}) is questionable although it gives reasonable results.
On the other hand, when taking into account the magnon damping phenomenologically this decoupling can be justified by the mode-selective effect of magnon damping.
Thus, when investigating parametrically pumped magnon gases, magnon damping should be accounted for.

\section{\label{sec:dynamics}
Collisionless magnon dynamics}
\begin{figure}[tbhp]
 \begin{minipage}{10pt}
  (a)
 \end{minipage}
 \begin{minipage}{0.92\linewidth}
  \includegraphics[clip=true,trim=50pt 180pt 45pt 200pt,width=\linewidth]{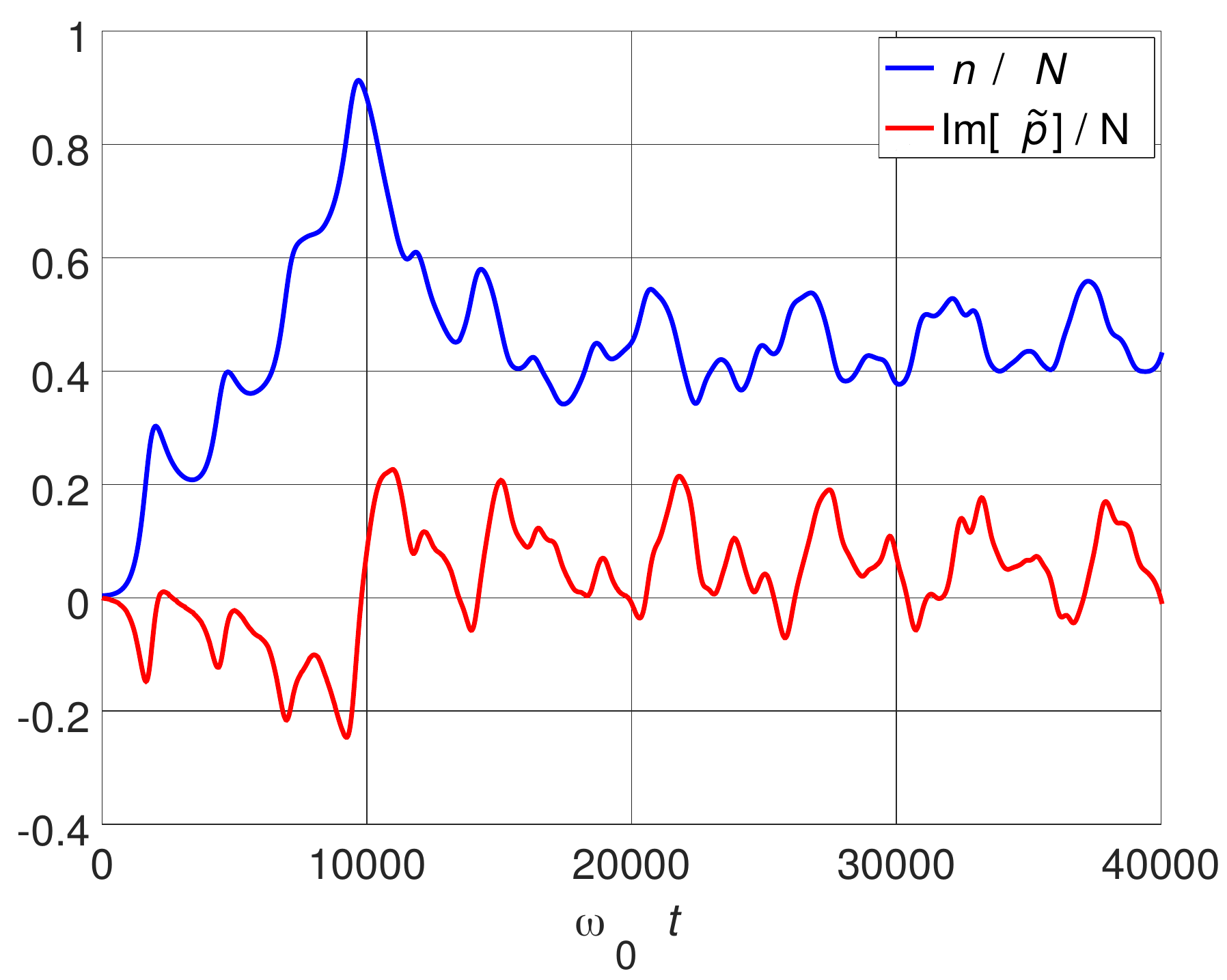}
 \end{minipage}
 \begin{minipage}{10pt}
  (b)
 \end{minipage}
 \begin{minipage}{0.92\linewidth}
  \includegraphics[clip=true,trim=50pt 180pt 45pt 200pt,width=\linewidth]{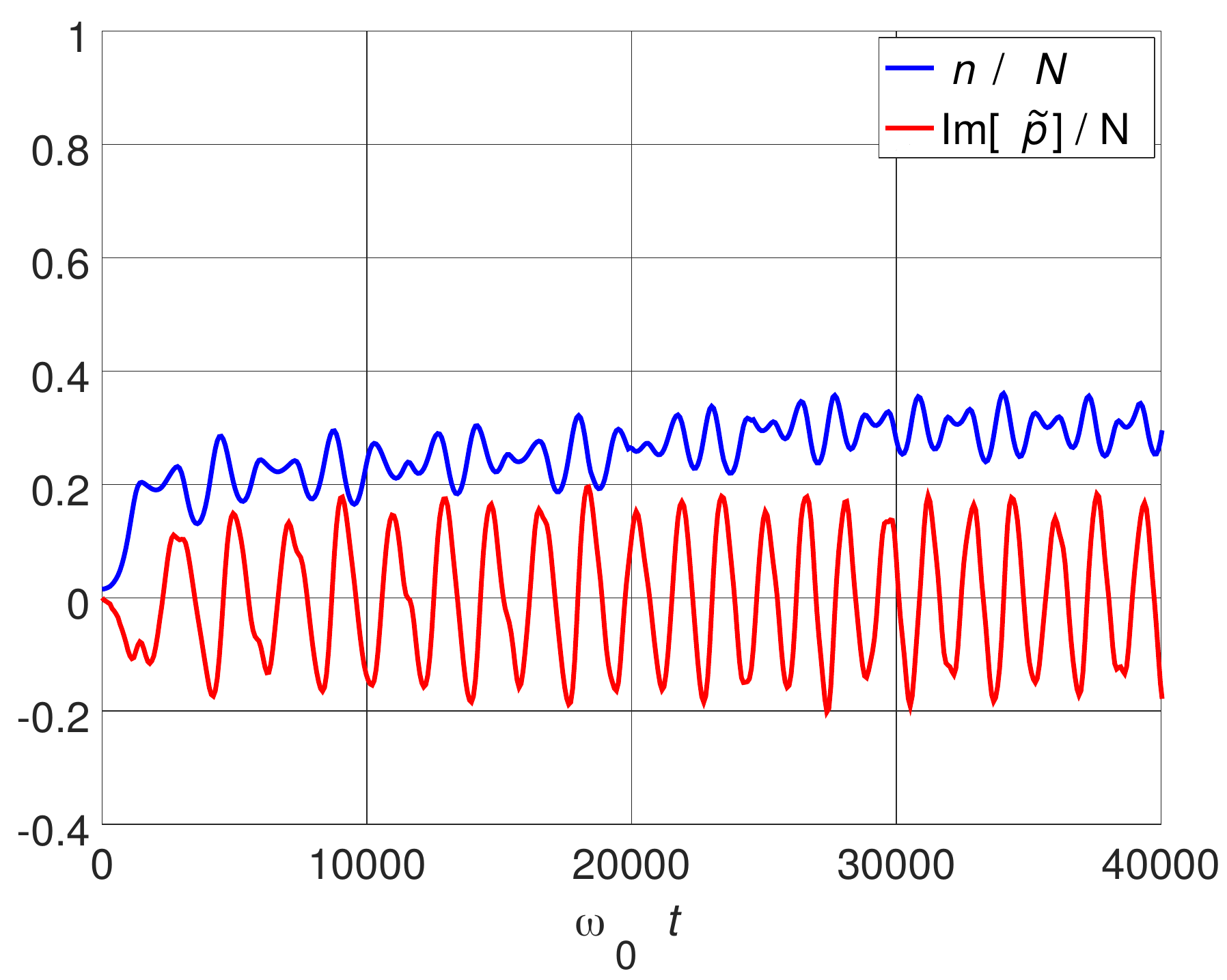}
\end{minipage}
 \begin{minipage}{10pt}
  (c)
 \end{minipage}
 \begin{minipage}{0.92\linewidth}
  \includegraphics[clip=true,trim=50pt 180pt 45pt 200pt,width=\linewidth]{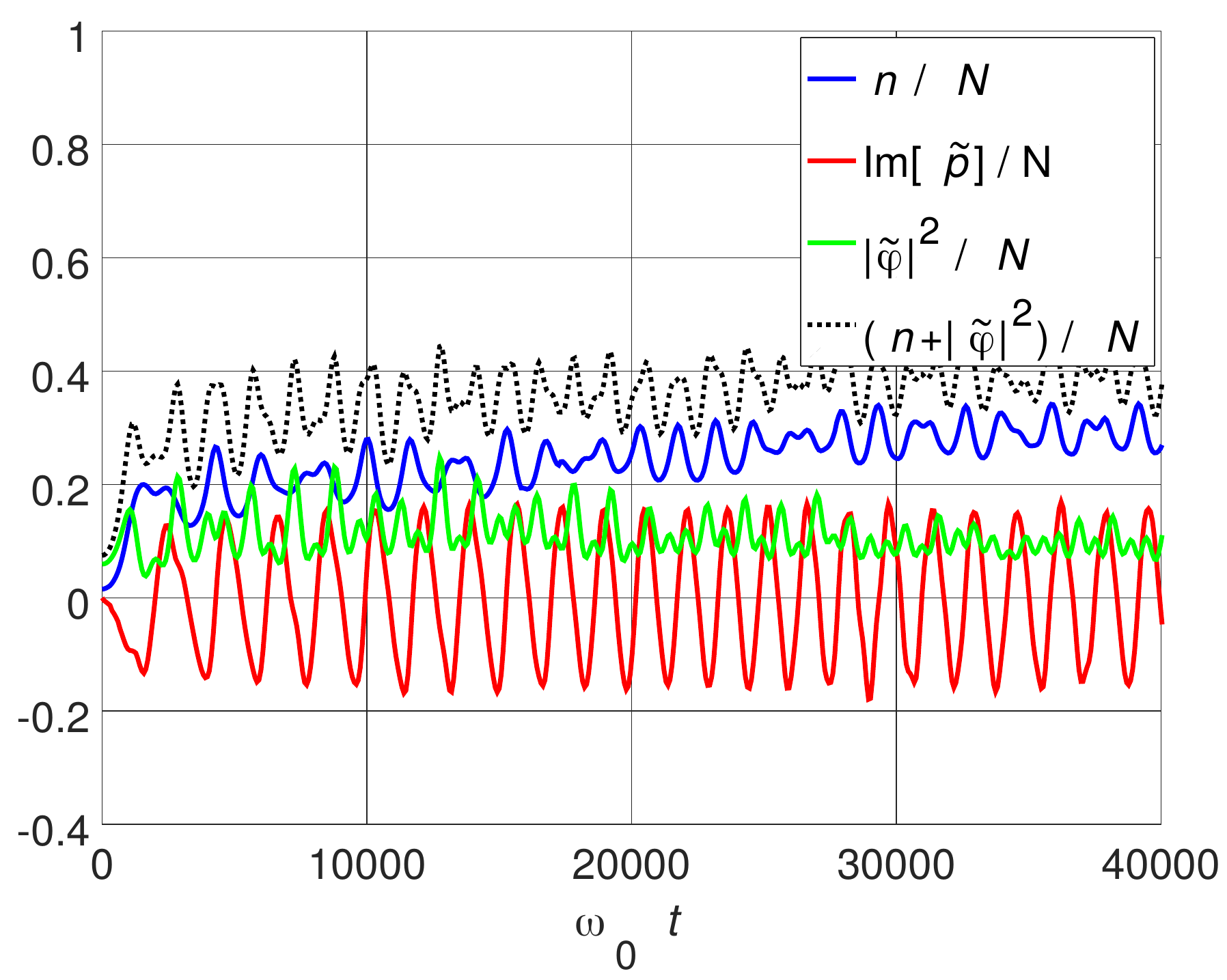}
 \end{minipage}
 \caption{Magnon density $n (t ) /N$ (blue) and imaginary part of the off-diagonal density 
${\rm{Im}} \tilde{p} (t )  /N$ (red) in the rotating reference frame obtained from the 
numerical solution of the collisionless kinetic equations 
 \eqref{eq:dgl_n0}--\eqref{eq:dgl_a0}  
at different levels of approximation
discussed in Sect.~\ref{sec:kinetic_eq}:
(a) conventional S-theory, (b) S-theory with mode coupling, 
(c) S-theory with mode coupling and magnon condensation; 
the green line represents $ \sum_{\bd{k}} |\tilde{\psi}_{\bd{k}} |^2 / N$.
The results are obtained with momentum mesh consisting of $N_\alpha = 30$ 
angles and $N_\varepsilon = 20$ energies as described in the text. 
The parameters are chosen as $H_0 = 800$ Oe, 
$H_1 = 50$ Oe, $\omega_0 = 13.857$ GHz, and $d=10\mu$m corresponding to $N=8080$.
}
 \label{fig:dyn}
\end{figure}

\begin{figure}[tbhp]
 \begin{minipage}{10pt}
  (a)
 \end{minipage}
 \begin{minipage}{0.92\linewidth}
  \includegraphics[clip=true,trim=20pt 180pt 45pt 200pt,width=\linewidth]{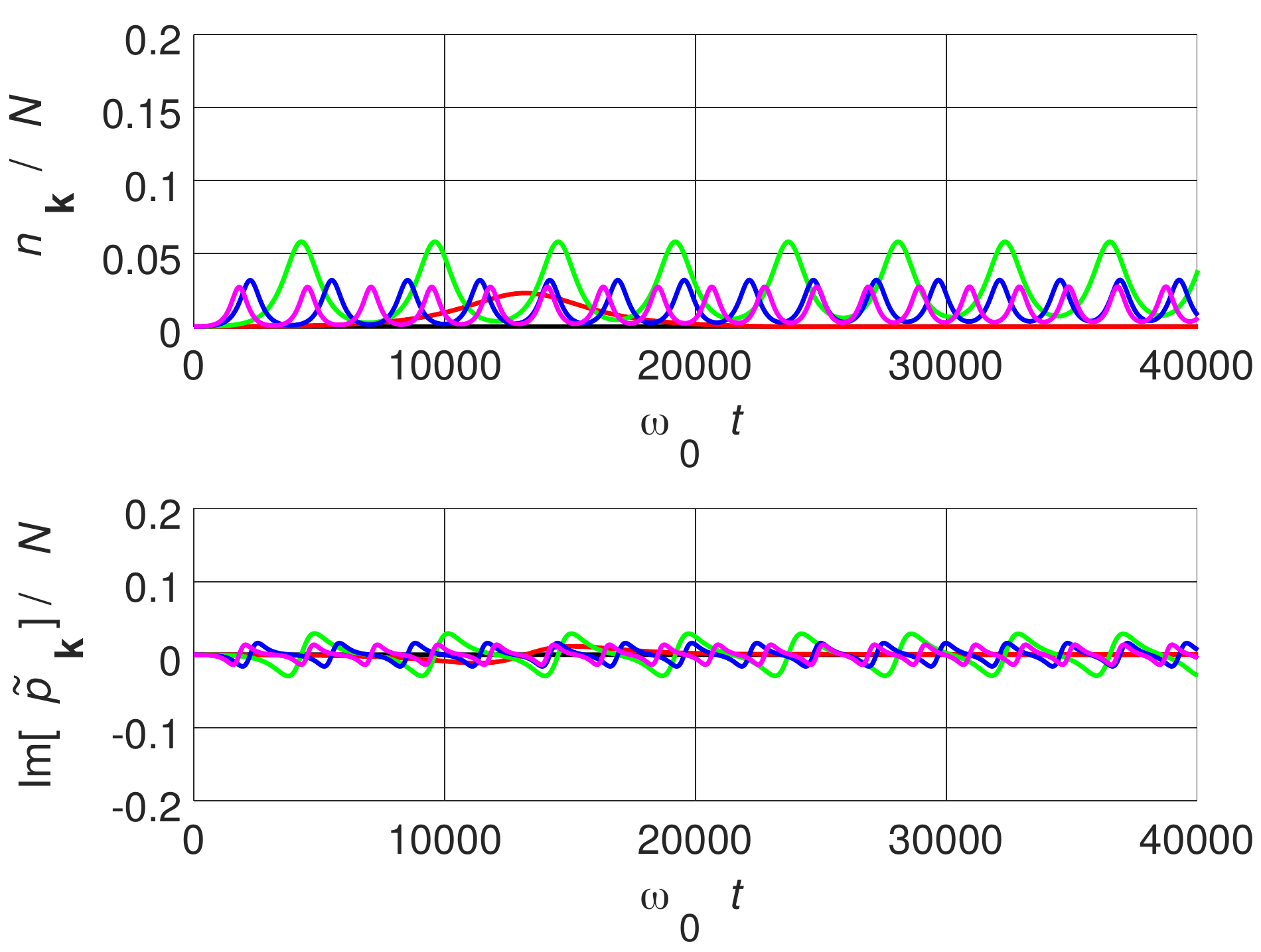}
 \end{minipage}
 \begin{minipage}{10pt}
  (b)
 \end{minipage}
 \begin{minipage}{0.92\linewidth}
  \includegraphics[clip=true,trim=20pt 180pt 45pt 200pt,width=\linewidth]{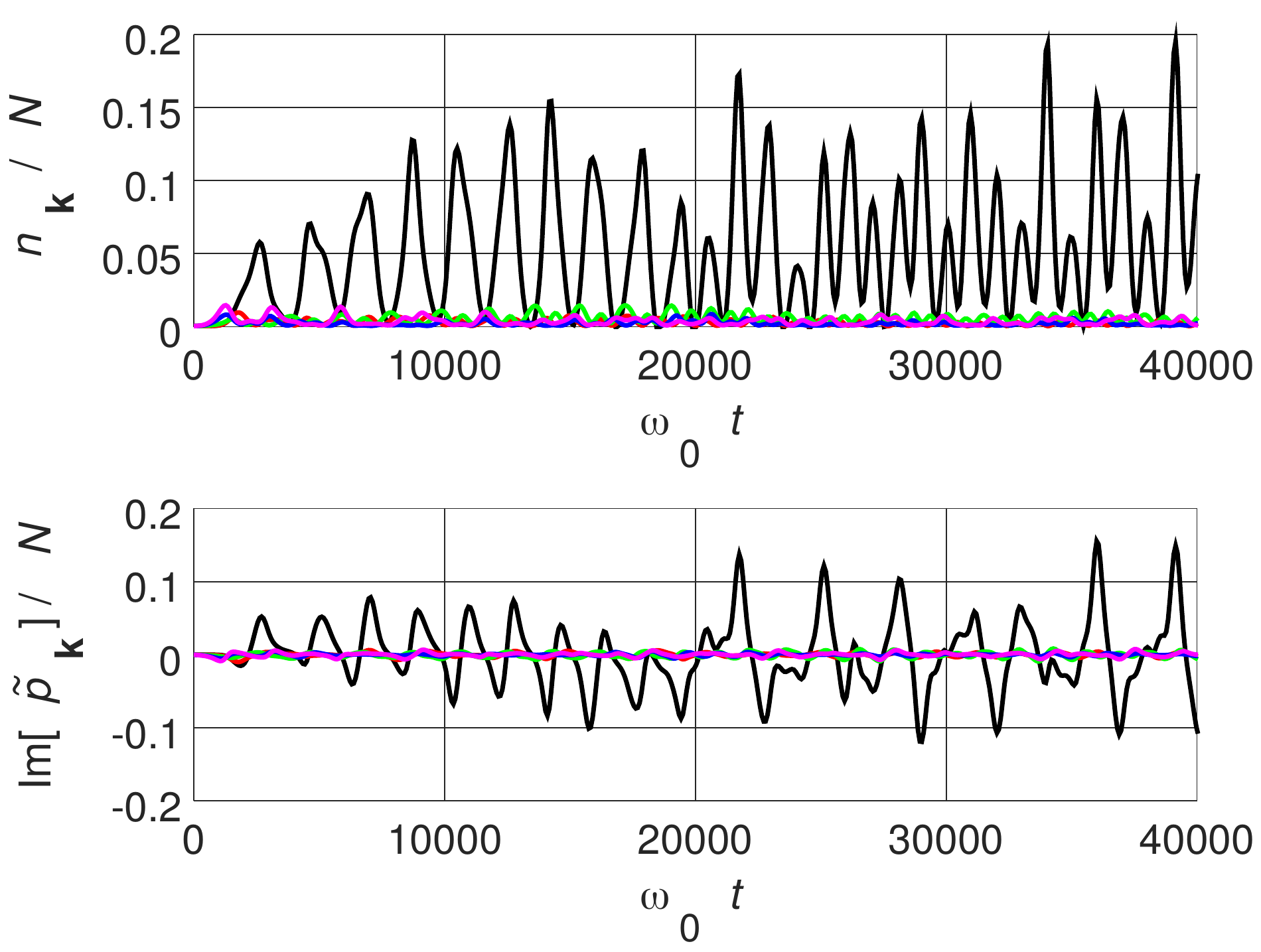}
\end{minipage}
 \begin{minipage}{10pt}
  (c)
 \end{minipage}
 \begin{minipage}{0.92\linewidth}
  \includegraphics[clip=true,trim=20pt 190pt 45pt 200pt,width=\linewidth]{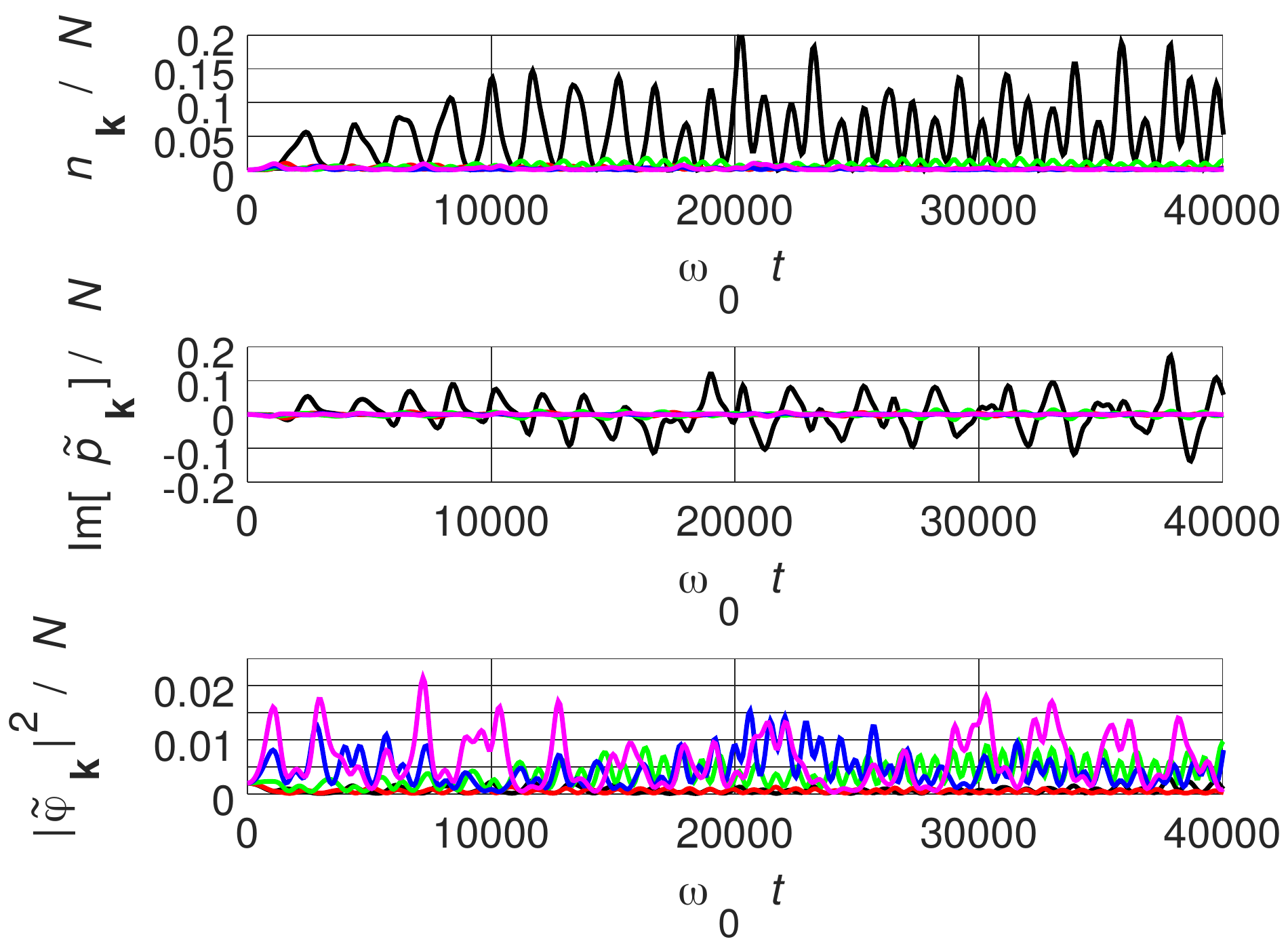}
 \end{minipage}
 \caption{
Time evolution of the diagonal distribution $n_\kv(t) / N$ (top) and the off-diagonal distribution function $\tilde{p}_\kv(t) / N$ (bottom) for different wavevectors satisfying the resonance condition $\varepsilon_\kv=\omega_0/2$.
We have chosen six different angles $\theta_\kv$ between $0$ and $\pi/2$ uniformly to be shown.
(a) within conventional S-theory, (b) S-theory with mode coupling, (c) S-theory with mode coupling and magnon condensation.
The parameters are the same as in Fig.~\ref{fig:dyn}.}
\label{fig:dyn_modes}
\end{figure}

\begin{figure}[tbhp]
 \begin{minipage}{10pt}
  (a)
 \end{minipage}
 \begin{minipage}{0.92\linewidth}
  \includegraphics[clip=true,trim=20pt 170pt 45pt 200pt,width=\linewidth]{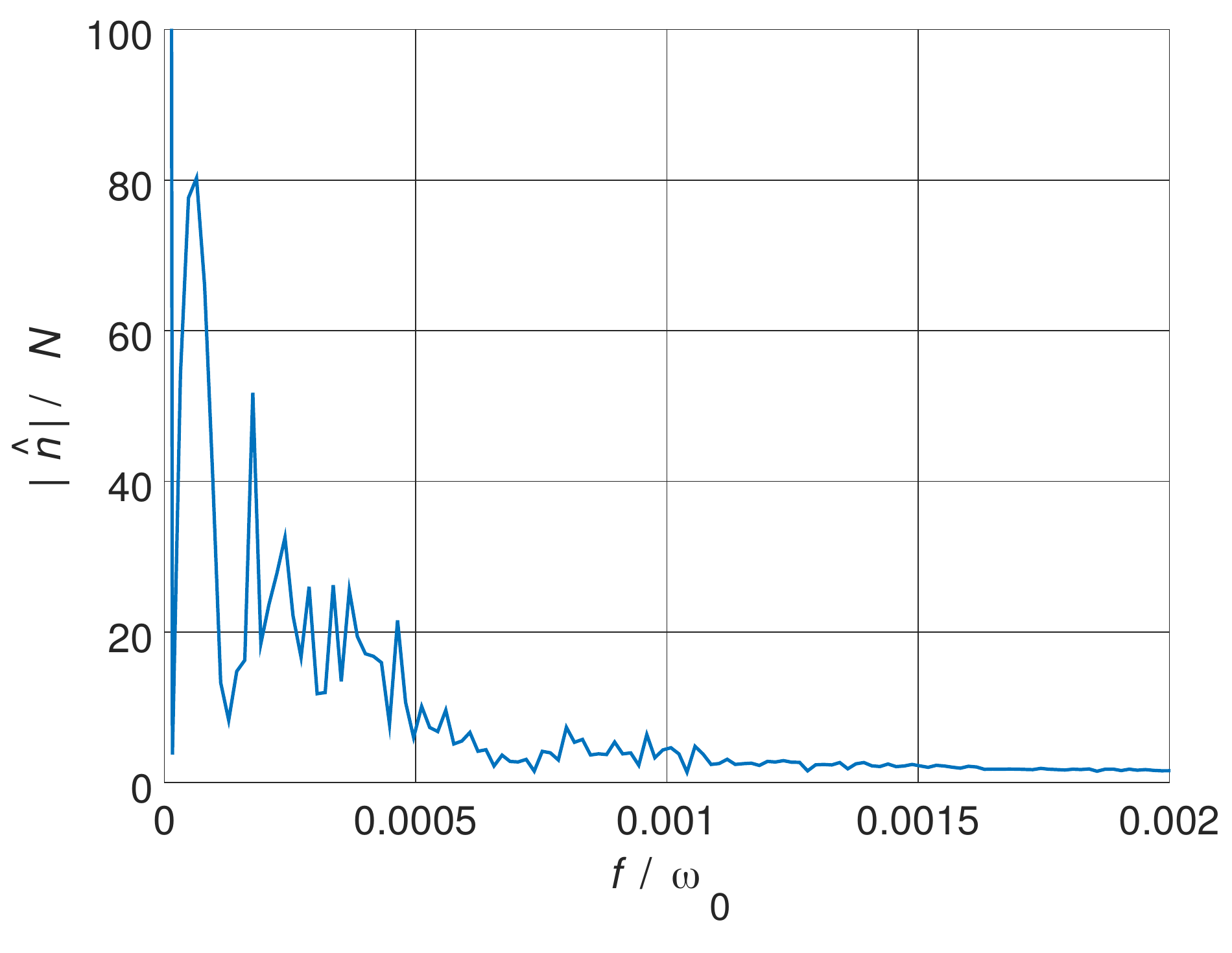}
 \end{minipage}
 \begin{minipage}{10pt}
  (b)
 \end{minipage}
 \begin{minipage}{0.92\linewidth}
  \includegraphics[clip=true,trim=20pt 170pt 45pt 200pt,width=\linewidth]{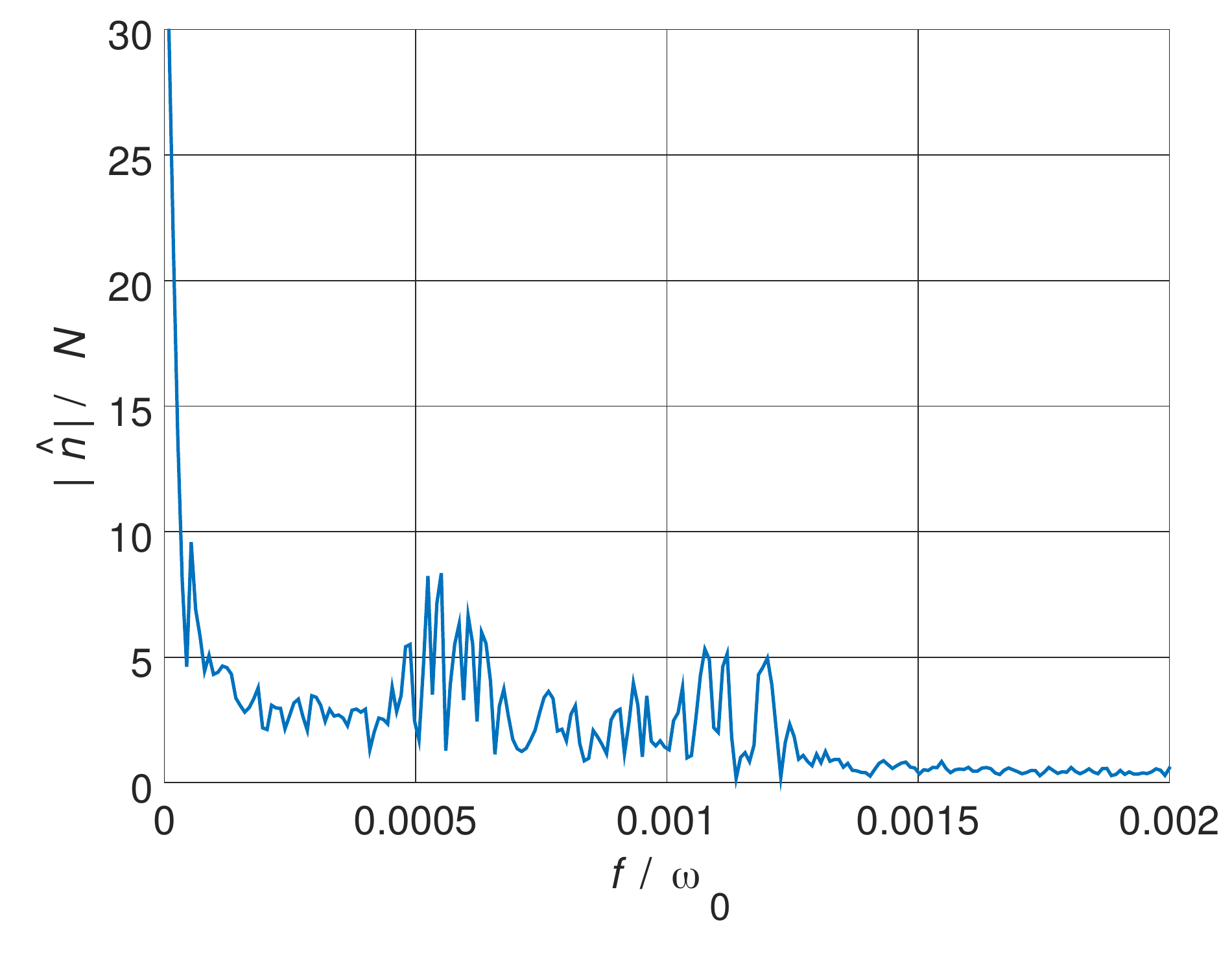}
 \end{minipage}
 \begin{minipage}{18pt}
  (c)
 \end{minipage}
 \begin{minipage}{0.92\linewidth}
  \includegraphics[clip=true,trim=20pt 170pt 45pt 200pt,width=\linewidth]{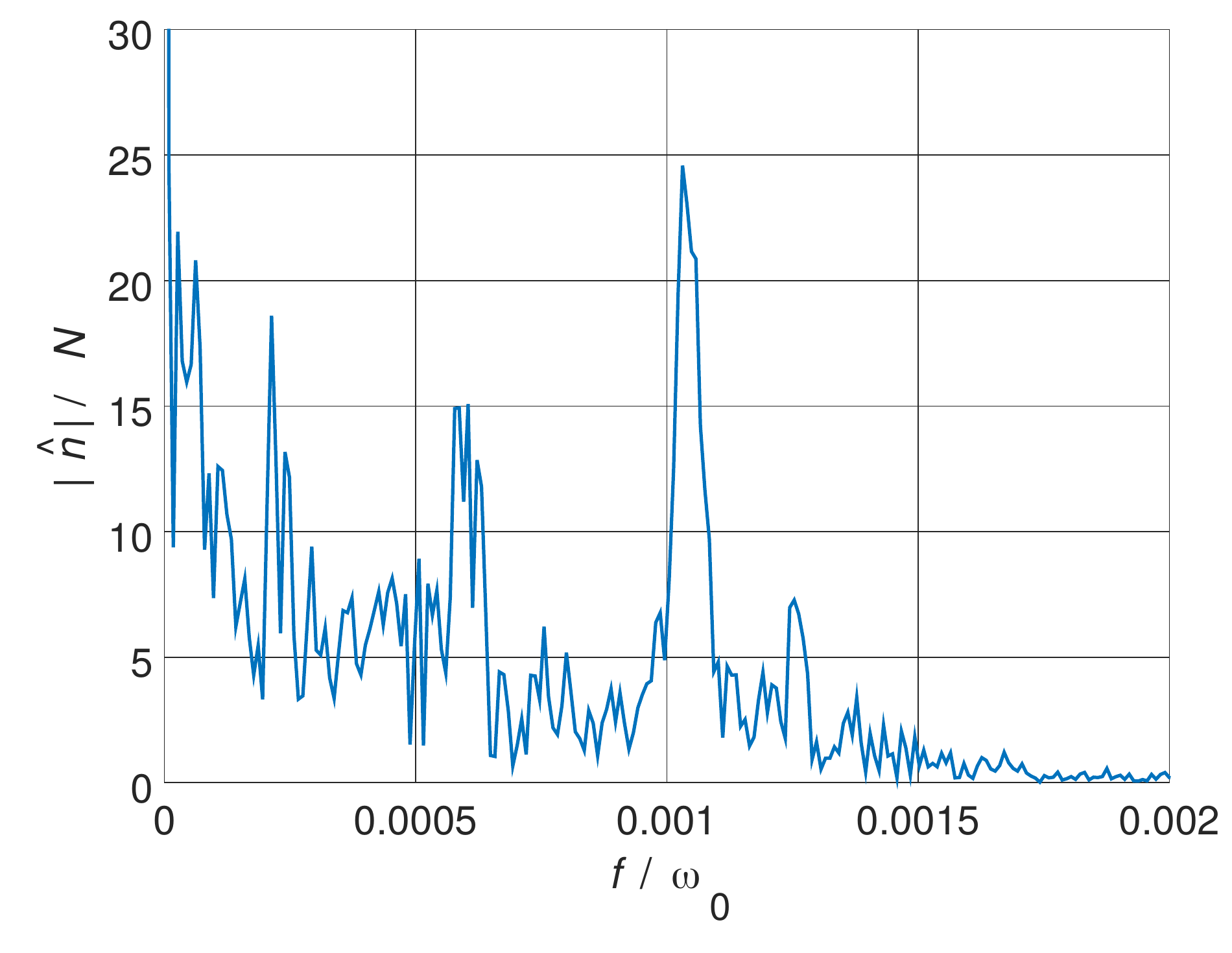}
 \end{minipage}
 \caption{The absolute value of the discrete Fourier transform $\hat n(f_l) = \sum_{j=0}^{N_t-1} e^{-2\pi i f_l j} n(t_j)$ of the magnon density $n(t)=\sum_{\bd{k}} n_{\bd{k}}(t)$ shown in Fig.~\ref{fig:dyn}, where $f_l = \frac{l}{N_t}$.
  (a) Within conventional S-theory, (b) S-theory with mode coupling, (c) S-theory with mode coupling and magnon condensation. In this case the Fourier transform of $n^c(t)+|\psi(t)|^2$ is shown.}
 \label{fig:ft}
\end{figure}

\fussy
In this section we present numerical results for the time evolution
of the magnon distribution obtained from the solution of
the collisionless kinetic equations \eqref{eq:dgl_n0}--\eqref{eq:dgl_a0} of modified S-theory with magnon condensation and mode coupling and compare the results with the predictions of conventional S-theory discussed in Sect.~\ref{sec:S-theory}.
For the numerical solution of the kinetic equations 
we discretize the momenta in the same way as above, shown in Fig.~\ref{fig:discretization}.
The total magnon number and off-diagonal occupation in the rotating reference frame are then approximated by
 \begin{subequations}
 \begin{eqnarray}
 n & = &  \sum_{\bd{k}} n_{\bd{k}}  \approx \sum_{ i=1}^{ N_{\varepsilon}} \sum_{ j =1}^{ N_{\alpha}}
 n_{ \bd{k}_{ij} }, \\
 \tilde{p} & = &  \sum_{\bd{k}} \tilde{p}_{\bd{k}}  
\approx \sum_{ i=1}^{ N_{\varepsilon}} \sum_{ j =1}^{ N_{\alpha}}
 \tilde{p}_{ \bd{k}_{ij} },
 \end{eqnarray}
 \end{subequations}
where $\bd{k}_{ij}$ is a wavevector in the center of the mesh-points.
Our numerical results for the total magnon density $n / N$ and the imaginary part of the
off-diagonal density $ {\rm Im} \tilde{p} / N$
are shown in Fig.~\ref{fig:dyn}, where we compare these quantities for the three different
versions of S-theory developed in Sect.~\ref{sec:kinetic_eq}.
The numerical results are obtained for a thin YIG film with thickness $d = 10 \, \mu$m, exposed to a static magnetic field $H_0 = 800$ Oe and oscillating field with amplitude
$H_1 = 50$ Oe and frequency $\omega_0 = 13.857$ GHz.
In Fig.~\ref{fig:dyn}~(a) we show the time evolution of the  
total magnon density $n / N$ and of ${\rm Im} \tilde{p} /N$  as predicted by the conventional
S-theory and mode decoupling as discussed in Sect.~\ref{sec:mode_decoupling}.
Because the damping is neglected, the magnon density does not approach a stationary limit but continues to oscillate around the value $n/N = 0.46$ which agrees with the stationary value
$n = \sum_{\bd{k}} n_{\bd{k}}$ obtained from Eq.~(\ref{eq:n_stat_S}).
Note that in this approximation modes with different wavevectors are completely decoupled.
In Fig.~\ref{fig:dyn_modes}~(a)
we show the time evolution of the occupation number of magnon modes for some representative
wavevectors.
Due to the mode-decoupling, the occupations oscillate with given 
characteristic frequencies and amplitudes; summing over all wavevectors we recover
the time evolution of the total densities shown in  Fig.~\ref{fig:dyn}. 
Fig.~\ref{fig:ft} shows the Fourier transform of the magnon density which has peaks for low frequencies.
As the amplitudes of the modes vary greatly, the time evolution of the total magnon density
is dominated by the modes satisfying the parametric resonance
condition  $\left|V_\kv\right| > \left|E_\kv\right|$. If we completely neglect
magnon-magnon interactions, then the magnon occupation grows exponentially in this regime; the Hartree-Fock correlations retained within S-theory cut off the exponential growth and eventually lead to an oscillatory behavior with time average given by the  stationary distribution derived in Eq.~(\ref{eq:n_stat_S}).

The above results rely on the mode-decoupling approximation 
\eqref{eq:sum_approx}. If we do not use this approximation but
retain all modes in integrals of the self-energy contributions to the 
renormalized magnon energy $\tilde{E}_{\bd{k}} ( t )$ and the pumping energy
$\tilde{V}_{\bd{k}} ( t )$ defined in Eq.~(\ref{eq:HFenergies}) of Sect.~\ref{sec:S-theory}, 
the time evolution looks rather different, as shown in Figs.~\ref{fig:dyn}~(b) and ~\ref{fig:dyn_modes}~(b).
The most striking difference is that now the time-dependence of the magnon density
is a superposition of oscillations with different frequencies, while the off-diagonal density
$p$ oscillates with fixed frequency and almost constant amplitude. 
Note also that $n (t ) /N$ now oscillates around $n/N \approx 0.30$,
which is significantly smaller than the corresponding value $n/N \approx 0.46$
obtained within the mode-decoupling approximation.
The time evolution of specific momentum modes shown in Fig.~\ref{fig:dyn_modes}~(b)
exhibits a rather complex behavior which
now does not even resemble its mode-decoupled counter-part 
in Fig.~\ref{fig:dyn_modes}~(a). This change is also visible in frequency space (Fig.~\ref{fig:ft} (b)); the peaks for low frequency become less dominant and higher frequencies emerge.

Finally, let us also take the magnon condensation into account as described in Sect.~\ref{sec:Smag}.
The result of the numerical solution of the coupled integro-differential equations
(\ref{eq:dgl_n0})--(\ref{eq:EVint}) is shown in
Figs.~\ref{fig:dyn}~(c) and~\ref{fig:dyn_modes}~(c).
The time evolution of the total magnon density looks very similar
to the corresponding time evolution without magnon condensation, with a slight\-ly smaller time-average 
$(n+|\tilde\psi|^2) / N \approx 0.29$. However, the
time evolution of representative modes shown in Fig.~\ref{fig:dyn_modes}~(c) is again different from the previous two cases. In general, the inclusion of
the finite expectation values of the magnon operators leads to faster oscillations 
involving a broader range of frequencies.

\section{\label{sec:conclusions}
Summary and conclusions}

In this work we have extended S-theory, which is a collisionless kinetic theory for pumped magnon gases, by including the coupling between different modes and the finite expectation values of the
magnon operators. The conventional equations of S-theory should then be complemented
by an additional equation of motion for the expectation values of the magnon operators which is analogous to the
Gross-Pitaevskii equation of a Bose gas.

We have numerically solved collisionless kinetic equations for diagonal- and off-diagonal 
distribution functions of magnons 
and have compared the resulting magnon dynamics using three different approximation schemes:
conventional S-theory without mode coupling, 
S-theory with mode coupling, and S-theory with mode coupling and mag\-non condensation. 
The time-averaged magnon density in the saturated regime has been found to have the 
same order of magnitude for all three cases. However, the different approximations lead to
a very different time-dependence of the occupation of 
representative magnon modes with given wavevectors: mode coupling
generates a more complex time evolution 
which still exhibits some periodic structures, 
which are destroyed if the
dynamics of the expectation values of the magnon operators is taken into account.

We have found numerically that a unique solution of the collisionless kinetic equations exists only in some re\-gimes of parameter space and have argued that the solution is unphysical.
We conclude that physically meaningful results for the stationary non-equilibrium state of pumped magnon gases can only be obtained if magnon damping is taken into account,
either by including the microscopic collision integrals or phenomenologically by adding a damping rate by hand.
As discussed in Sect.~\ref{sec:stat}, the latter is usually done in order to justify the mode-decoupling approximation within conventional S-theory.
While such a procedure can  be physically motivated, it is formally not satisfying.
We have recently made some progress in this direction \cite{footnoteyig} by
solving the kinetic equations for the magnon distribution in YIG including
the microscopic collision integral arising from the cubic magnon vertices.

\section*{Author contribution statement}
Viktor Hahn carried out all numerical and analytical calculations; Peter Kopietz has checked most of the analytical calculations. Both authors have discussed the results and have contributed to writing the paper.

\begin{appendix}

\section*{APPENDIX: HAMILTONIAN FOR
PUMPED MAGNONS IN YIG}
\setcounter{equation}{0}
\renewcommand{\theequation}{A\arabic{equation}}
 \label{sec:ward}
To make this work self-contained, 
we outline here the main technical steps in the derivation of the magnon Hamiltonian
given in Eqs.~(\ref{eq:H_2}) and (\ref{eq:H4rot}) from the effective spin 
Hamiltonian (\ref {eq:eff_H}), see also Refs.~\cite{Hick10,Kreisel09}.
As usual,  we express the components
of the spin operators $\bd{S}_i$ 
in terms of boson annihilation and creation operators $b_i$ and $b^{\dagger}_i$
using the Holstein-Primakoff transformation \cite{Holstein40} 
and expand the effective boson Hamiltonian in powers of $1/S$, see Eq.\eqref{eq:H_1/S}.
Transforming to momentum space,
\begin{equation}
 b_i = \frac{1}{\sqrt{N}} \sum\limits_\kv \text{e}^{i \kv \cdot \bd{r}_i} b_\kv,
\end{equation}
the quadratic part $\mathcal{H}_2(t)$ of the effective boson Hamiltonian can be written as~\cite{Hick10}
\begin{eqnarray}
 \mathcal{H}_2(t)&=&\sum\limits_\kv\left[A_\kv b^\dagger_\kv b_\kv + \frac{B_\kv}{2}\left(b^\dagger_\kv b^\dagger_{-\kv}+b_{-\kv}b_\kv\right)\right] \nonumber\\*
 &&+h_1\cos\left(\omega_0t\right)\sum\limits_\kv b^\dagger_\kv b_\kv ,
 \label{eq:H2_ft}
\end{eqnarray}
where
\begin{eqnarray}
A_\kv &=& h_0 + S \left(J_{\bd{0}} - J_\kv\right) + S \left[D^{zz}_{\bd{0}} - \frac{1}{2}\left(D^{xx}_\kv + D^{yy}_\kv\right)\right], \nonumber\\
\label{eq:Akdef} \\
B_\kv &=& -\frac{S}{2} \left[D^{xx}_\kv - 2 i D^{xy}_\kv - D^{yy}_\kv\right],
 \label{eq:Bkdef}
\end{eqnarray}
and $J_\kv$ and $D^{\alpha \beta}_{\kv}$ are the Fourier transforms of the exchange and dipolar couplings,
\begin{eqnarray}
J_\kv &=& \sum\limits_i \text{e}^{-i \kv \cdot \bd{r}_{i j}} J_{i j}, \\*
D^{\alpha \beta}_\kv &=& \sum\limits_i \text{e}^{-i \kv \cdot \bd{r}_{i j}} D^{\alpha \beta}_{i j}.
\end{eqnarray}
The cubic and quartic parts of the Hamiltonian read~\cite{Hick10}
\begin{eqnarray}
 \mathcal{H}_3 &=& \frac{1}{\sqrt{N}} \sum\limits_{\kv_1, \kv_2, \kv_3} \delta_{\kv_1+\kv_2+\kv_3, 0} \frac{1}{2!} \left[\Gamma^{\bar{b}bb}_{1; 2, 3} b^\dagger_{-1} b_2 b_3 \right.\nonumber\\
 &&\left.+ \Gamma^{\bar{b}\bar{b}b}_{1, 2; 3} b^\dagger_{-1} b^\dagger_{-2} b_3\right] ,\\
\mathcal{H}_4 &=& \frac{1}{N} \sum\limits_{\kv_1, \dots, \kv_4} \delta_{\kv_1+\kv_2+\kv_3+\kv_4, 0} \left[\frac{1}{\left(2!\right)^2} \Gamma^{\bar{b}\bar{b}bb}_{1, 2; 3, 4} b^\dagger_{-1} b^\dagger_{-2} b_3 b_4 \right.\nonumber\\*
&&\left.+ \frac{1}{3!} \Gamma^{\bar{b}bbb}_{1; 2, 3, 4} b^\dagger_{-1} b_2 b_3 b_4  + \frac{1}{3!} \Gamma^{\bar{b}\bar{b}\bar{b}b}_{1, 2, 3; 4} b^\dagger_{-1} b^\dagger_{-2} b^\dagger_{-3} b_4\right], \nonumber\\\label{eq:H4_ft}
\end{eqnarray}
where the cubic and quartic vertices are given by
\begin{subequations}
\begin{eqnarray}
\Gamma^{\bar{b}bb}_{1; 2, 3} &=& \sqrt{\frac{S}{2}} \left[D^{zy}_{\kv_2} - i D^{zx}_{\kv_2} + D^{zy}_{\kv_3} - i D^{zx}_{\kv_3} \right.\nonumber\\
 &&\left.+ \frac{1}{2} \left(D^{zy}_{\mathbf{0}} - i D^{zx}_{\mathbf{0}}\right)\right] , \\
 \Gamma^{\bar{b}\bar{b}b}_{1, 2; 3} &=& \left(\Gamma^{\bar{b}bb}_{3; 2, 1}\right)^* ,\\
\Gamma^{\bar{b}\bar{b}bb}_{1, 2; 3, 4} &=& -\frac{1}{2}\left[J_{\kv_1+\kv_3} + J_{\kv_2+\kv_3} + J_{\kv_1+\kv_4} + J_{\kv_2+\kv_4} \right.\nonumber\\*
&&\left.+ D^{zz}_{\kv_1+\kv_3} + D^{zz}_{\kv_2+\kv_3} + D^{zz}_{\kv_1+\kv_4} + D^{zz}_{\kv_2+\kv_4}\right.\nonumber\\*
&&\left. - \sum\limits_{i=1}^4 \left(J_{\kv_i} - 2 D^{zz}_{\kv_i}\right)\right], \\
\Gamma^{\bar{b}bbb}_{1; 2, 3, 4} &=& \frac{1}{4} \left[D^{xx}_{\kv_2} - 2 i D^{xy}_{\kv_2} - D^{yy}_{\kv_2} + D^{xx}_{\kv_3}-2 i D^{xy}_{\kv_3} - D^{yy}_{\kv_3} \right.\nonumber\\*
&&\left. + D^{xx}_{\kv_4} - 2 i D^{xy}_{\kv_4} - D^{yy}_{\kv_4}\right], \\*
\Gamma^{\bar{b}\bar{b}\bar{b}b}_{1, 2, 3; 4} &=& \left(\Gamma^{\bar{b}bbb}_{4; 1, 2, 3}\right)^* .
\end{eqnarray}
\end{subequations}
To diagonalize the time-independent part of ${\cal{H}}_2 ( t )$ we use
a canonical transformation
 \begin{equation}
 \left( \begin{array}{c} b_{\bd{k}} \\ b^{\dagger}_{ - \bd{k}} \end{array} \right)
 = \left( \begin{array}{cc} u_{\bd{k}} & - v_{\bd{k}} \\
 - v_{\bf{k}}^{\ast} & u_{\bd{k}} \end{array} \right)
\left( \begin{array}{c} a_{\bd{k}} \\ a^{\dagger}_{ - \bd{k}} \end{array} \right),
 \end{equation}
where
\begin{subequations}
\begin{eqnarray}
u_\kv &=& \sqrt{\frac{A_\kv + \varepsilon_\kv}{2 \varepsilon_\kv}}, \\
v_\kv &=& \frac{B_\kv}{|B_\kv|} \sqrt{\frac{A_\kv - \varepsilon_\kv}{2 \varepsilon_\kv}},
\end{eqnarray}
\end{subequations}
and the magnon dispersion is given by
\begin{equation}
\varepsilon_\kv = \sqrt{A_\kv^2 - |B_\kv|^2} .
\end{equation}
Due to the time dependence of the last term in Eq.~\eqref{eq:H2_ft}, the quadratic part of the 
Hamiltonian has also off-diagonal terms,
\begin{eqnarray}
 \mathcal{H}_2(t) &=& \sum\limits_\kv \left[\varepsilon_\kv a^\dagger_\kv a_\kv + \frac{\varepsilon_\kv - A_\kv}{2} \right.\nonumber\\*
 &&\left.\hspace{-14pt}+ h_1 \cos\left(\omega_0 t\right) \left(\frac{A_\kv}{\varepsilon_\kv} a^\dagger_\kv a_\kv - \frac{\varepsilon_\kv - A_\kv}{2 \varepsilon_\kv}\right)\right] \nonumber\\*
 &&\hspace{-14pt}+ \sum\limits_\kv \left[V_\kv \cos\left(\omega_0 t\right) a^\dagger_\kv a^\dagger_{-\kv} + V_\kv^* \cos\left(\omega_0 t\right) a_{-\kv}a_\kv\right], \nonumber\\*
 \label{eq:H2_r}
\end{eqnarray}
where
\begin{equation}
V_\kv = -\frac{h_1 B_\kv}{2 \varepsilon_\kv} .
\end{equation}
At this point we transform to the rotating reference frame 
using another canonical transformation given in Eq.~(\ref{eq:rotref}).
In rotating-wave approximation all terms which still exhibit an explicit time-dependence in the rotating reference frame are neglected so that
we arrive at the time-dependent quadratic Hamiltonian
in Eq.~\eqref{eq:H2_t}.

In terms of the magnon operators $\tilde{a}_{\bd{k}}$ and $\tilde{a}^{\dagger}_{\bd{k}}$ in the rotating reference frame
the cubic and quartic parts of the Hamiltonian, $\tilde{\mathcal{H}}_3$ and $\tilde{\mathcal{H}}_4$, 
are~\cite{Hick10}
\begin{eqnarray}
 \tilde{\mathcal{H}}_3 &=& \frac{1}{\sqrt{N}} \sum\limits_{\kv_1, \kv_2, \kv_3} \delta_{\kv_1+\kv_2+\kv_3, 0} \left[\frac{1}{2} \Gamma^{\bar{a}aa}_{1; 2, 3} e^{-i \omega_0 t / 2} \tilde{a}^\dagger_{-1} \tilde{a}_2 \tilde{a}_3 \right.\nonumber\\*
 &&\left. + \frac{1}{2} \Gamma^{\bar{a}\bar{a}a}_{1, 2; 3} e^{i \omega_0 t / 2} \tilde{a}^\dagger_{-1} \tilde{a}^\dagger_{-2} \tilde{a}_3 \right.\nonumber\\
 &&\left.+ \frac{1}{3!} \Gamma^{aaa}_{1, 2, 3} e^{-3 i \omega_0 t/2} \tilde{a}_1 \tilde{a}_2 \tilde{a}_3 \right.\nonumber\\*
 &&\left. + \frac{1}{3!} \Gamma^{\bar{a}\bar{a}\bar{a}}_{1, 2, 3} e^{3 i \omega_0 t/2} \tilde{a}^\dagger_{-1} \tilde{a}^\dagger_{-2} \tilde{a}^\dagger_{-3}\right],\\
 \tilde{\mathcal{H}}_4 &=& \frac{1}{N} \sum\limits_{\kv_1, \dots, \kv_4} \delta_{\kv_1+\dots+\kv_4, 0} \left[\frac{1}{\left(2!\right)^2}\Gamma^{\bar{a}\bar{a}aa}_{1, 2; 3, 4} \tilde{a}^\dagger_{-1} \tilde{a}^\dagger_{-2} \tilde{a}_3 \tilde{a}_4 \right.\nonumber\\*
 &&\left. + \frac{1}{3!} \Gamma^{\bar{a}aaa}_{1; 2, 3, 4} e^{-i \omega_0 t} \tilde{a}^\dagger_{-1} \tilde{a}_2 \tilde{a}_3 \tilde{a}_4 \right.\nonumber\\*
 &&\left. + \frac{1}{3!} \Gamma^{\bar{a}\bar{a}\bar{a}a}_{1, 2, 3; 4} e^{i \omega_0 t} \tilde{a}^\dagger_{-1} \tilde{a}^\dagger_{-2} \tilde{a}^\dagger_{-3} \tilde{a}_4 \right.\nonumber\\*
 &&\left. + \frac{1}{4!} \Gamma^{aaaa}_{1, 2, 3, 4} e^{-2i \omega_0 t} \tilde{a}_1 \tilde{a}_2 \tilde{a}_3 \tilde{a}_4 \right.\nonumber\\*
 &&\left. + \frac{1}{4!} \Gamma^{\bar{a}\bar{a}\bar{a}\bar{a}}_{1, 2, 3, 4} e^{2i \omega_0 t} \tilde{a}^\dagger_{-1} \tilde{a}^\dagger_{-2} \tilde{a}^\dagger_{-3} \tilde{a}^\dagger_{-4} \right] ,
\end{eqnarray}
with cubic vertices
\begin{subequations}
 \begin{eqnarray}
  \Gamma^{aaa}_{1, 2, 3} &=& - \Gamma^{\bar{b}bb}_{1; 2, 3} v_1 u_2 u_3 - \Gamma^{\bar{b}bb}_{2; 1, 3} v_2 u_1 u_3 - \Gamma^{\bar{b}bb}_{3; 1, 2} v_3 u_1 u_3 \nonumber\\*
  && + \Gamma^{\bar{b}\bar{b}b}_{1, 2; 3} v_1 v_2 u_3 + \Gamma^{\bar{b}\bar{b}b}_{2, 3; 1} v_2 v_3 u_1 + \Gamma^{\bar{b}\bar{b}b}_{1, 3; 2} v_1 v_3 u_2 , \nonumber\\*\\
 \Gamma^{\bar{a}aa}_{1; 2, 3} &=& \Gamma^{\bar{b}bb}_{1; 2, 3} u_1 u_2 u_3 + \Gamma^{\bar{b}bb}_{2; 1, 3} v_1 v_2 u_3 + \Gamma^{\bar{b}bb}_{3; 1, 2} v_1 v_3 u_2 \nonumber\\*
 && - \Gamma^{\bar{b}\bar{b}b}_{3, 2; 1} v_3 v_2 v_1 - \Gamma^{\bar{b}\bar{b}b}_{1, 2; 3} v_2 u_1 u_3 - \Gamma^{\bar{b}\bar{b}b}_{1, 3; 2} v_3 u_1 u_2 , \nonumber\\*\\
 \Gamma^{\bar{a}\bar{a}a}_{1, 2; 3} &=& \left(\Gamma^{\bar{a}aa}_{3; 2, 1}\right)^* , \\
 \Gamma^{\bar{a}\bar{a}\bar{a}}_{1, 2, 3} &=& \left(\Gamma^{aaa}_{1, 2, 3}\right)^* ,
 \end{eqnarray}
\end{subequations}
and quartic vertices
\begin{subequations}
\begin{eqnarray}
 \Gamma^{aaaa}_{1, 2, 3, 4} &=& \Gamma^{\bar{b}\bar{b}bb}_{1, 2; 3, 4} u_1 u_2 v_3 v_4 + \Gamma^{\bar{b}\bar{b}bb}_{1, 3; 2, 4} u_1 u_3 v_2 v_4 \nonumber\\*
 && + \Gamma^{\bar{b}\bar{b}bb}_{1, 4; 2, 3} u_1 u_4 v_2 v_3 + \Gamma^{\bar{b}\bar{b}bb}_{2, 3; 1, 4} u_2 u_3 v_1 v_4 \nonumber\\*
 && + \Gamma^{\bar{b}\bar{b}bb}_{2, 4; 1, 3} u_2 u_4 v_1 v_3 + \Gamma^{\bar{b}\bar{b}bb}_{3, 4; 1, 2} u_3 u_4 v_1 v_2 \nonumber\\*
 && - \Gamma^{\bar{b}bbb}_{4; 1, 2, 3} u_1 u_2 u_3 v_4 - \Gamma^{\bar{b}bbb}_{3; 1, 2, 4} u_1 u_2 u_4 v_3 \nonumber\\*
 && - \Gamma^{\bar{b}bbb}_{2; 1, 3, 4} u_1 u_3 u_4 v_2 - \Gamma^{\bar{b}bbb}_{1; 2, 3, 4} u_2 u_3 u_4 v_1 \nonumber\\*
 && - \Gamma^{\bar{b}\bar{b}\bar{b}b}_{2, 3, 4; 1} u_1 v_2 v_3 v_4 - \Gamma^{\bar{b}\bar{b}\bar{b}b}_{1, 3, 4; 2} u_2 v_1 v_3 v_4 \nonumber\\*
 && - \Gamma^{\bar{b}\bar{b}\bar{b}b}_{1, 2, 4; 3} u_3 v_1 v_2 v_4 - \Gamma^{\bar{b}\bar{b}\bar{b}b}_{1, 2, 3; 4} u_4 v_1 v_2 v_3, \nonumber\\*\label{eq:vertices1}\\
 \Gamma^{\bar{a}aaa}_{1; 2, 3, 4} &=& -\Gamma^{\bar{b}\bar{b}bb}_{2, 1; 3, 4} u_2 v_1 v_3 v_4 - \Gamma^{\bar{b}\bar{b}bb}_{3, 1; 2, 4} u_3 v_1 v_2 v_4 \nonumber\\*
 &&- \Gamma^{\bar{b}\bar{b}bb}_{4, 1; 2, 3} u_4 v_1 v_2 v_3 - \Gamma^{\bar{b}\bar{b}bb}_{2, 3; 1, 4} u_2 u_3 u_1 v_4 \nonumber\\*
 && - \Gamma^{\bar{b}\bar{b}bb}_{2, 4; 1, 3} u_2 u_4 u_1 v_3 - \Gamma^{\bar{b}\bar{b}bb}_{3, 4; 1, 2} u_3 u_4 u_1 v_2 \nonumber\\*
 && + \Gamma^{\bar{b}bbb}_{1; 2, 3, 4} u_1 u_2 u_3 u_4 + \Gamma^{\bar{b}bbb}_{4; 3, 2, 1} u_3 u_2 v_1 v_4 \nonumber\\*
 && + \Gamma^{\bar{b}bbb}_{3; 4, 2, 1} u_4 u_2 v_1 v_3 + \Gamma^{\bar{b}bbb}_{2; 4, 3, 1} u_4 u_3 v_1 v_2 \nonumber\\*
 && + \Gamma^{\bar{b}\bar{b}\bar{b}b}_{1, 2, 3; 4} u_4 u_1 v_2 v_3 + \Gamma^{\bar{b}\bar{b}\bar{b}b}_{1, 2, 4; 3} u_3 u_1 v_2 v_4 \nonumber\\*
 && + \Gamma^{\bar{b}\bar{b}\bar{b}b}_{1, 3, 4; 2} u_2 u_1 v_3 v_4 + \Gamma^{\bar{b}\bar{b}\bar{b}b}_{4, 3, 2; 1} v_4 v_2 v_3 v_1 , \nonumber\\*\\
 \Gamma^{\bar{a}\bar{a}aa}_{1, 2; 3, 4} &=& \Gamma^{\bar{b}\bar{b}bb}_{1, 2; 3, 4} u_1 u_2 u_3 u_4 + \Gamma^{\bar{b}\bar{b}bb}_{1, 3; 4, 2} u_1 u_4 v_3 v_2 \nonumber\\*
 && + \Gamma^{\bar{b}\bar{b}bb}_{1, 4; 3, 2} u_1 u_3 v_4 v_2 + \Gamma^{\bar{b}\bar{b}bb}_{2, 3; 4, 1} u_2 u_4 v_3 v_1 \nonumber\\*
 && + \Gamma^{\bar{b}\bar{b}bb}_{2, 4; 3, 1} u_2 u_3 v_4 v_1 + \Gamma^{\bar{b}\bar{b}bb}_{3, 4; 2, 1} v_1 v_2 v_3 v_4 \nonumber\\*
 && - \Gamma^{\bar{b}bbb}_{4; 3, 2, 1} u_3 v_2 v_1 v_4 - \Gamma^{\bar{b}bbb}_{3; 4, 2, 1} u_4 v_2 v_1 v_3 \nonumber\\*
 && - \Gamma^{\bar{b}bbb}_{2; 3, 4, 1} u_2 u_3 u_4 v_1 - \Gamma^{\bar{b}bbb}_{1; 3, 4, 2} u_1 u_3 u_4 v_2 \nonumber\\*
 && - \Gamma^{\bar{b}\bar{b}\bar{b}b}_{2, 3, 4; 1} u_2 v_3 v_4 v_1 - \Gamma^{\bar{b}\bar{b}\bar{b}b}_{1, 3, 4; 2} u_1 v_3 v_4 v_2 \nonumber\\*
 && - \Gamma^{\bar{b}\bar{b}\bar{b}b}_{1, 2, 4; 3} u_1 u_2 u_3 v_4 - \Gamma^{\bar{b}\bar{b}\bar{b}b}_{1, 2, 3; 4} u_1 u_2 u_4 v_3 , \nonumber\\*\\
 \Gamma^{\bar{a}\bar{a}\bar{a}\bar{a}}_{1, 2, 3, 4} &=& \Gamma^{aaaa}_{1, 2, 3, 4}, \\
 \Gamma^{\bar{a}\bar{a}\bar{a}a}_{1, 2, 3; 4} &=& \left(\Gamma^{\bar{a}aaa}_{4; 3, 2, 1}\right)^*. \label{eq:vertices5}
\end{eqnarray}
\end{subequations}

Finally, let us specify the coefficients $A_\kv$, $B_\kv$, and $D^{\alpha\beta}_\kv$
for a thin film of YIG. We assume that the thickness $d$ in $x$-direction is small compared to the extensions in $y$- and $z$-direction. We are only interested in the dispersion of the lowest magnon band and therefore use an effective in-plane Hamiltonian to derive the dispersion of the lowest magnon band.
We use the uniform mode approximation ignoring the fact that the system is not translationally invariant in the $x$-direction and replace the transverse mode by plane waves. This is valid for periodic boundary conditions in all directions \cite{Kreisel09}.
The coefficients $A_\kv$ and $B_\kv$ defined in  
Eqs.~(\ref{eq:Akdef}) and (\ref{eq:Bkdef})
can then be written as
\begin{eqnarray}
A_\kv &=& h_0 + J S \left[4 - 2 \cos\left(k_y a\right) - 2 \cos\left(k_z a\right)\right] \nonumber\\*
&& - \frac{S}{2} \left(D^{xx}_\kv + D^{yy}_\kv\right) + \frac{\Delta}{3},
\label{eq:Ak}\\
B_\kv &=& - \frac{S}{2} \left(D^{xx}_\kv - D^{yy}_\kv\right),
\end{eqnarray}
where the Fourier transforms of the dipole matrix elements are~\cite{Kreisel09}
\begin{subequations}
\begin{eqnarray}
D^{xx}_\kv &=& \frac{4 \pi \mu^2}{a^3} \left[\frac{1}{3} - f_\kv\right], \label{eq:Dxx}\\
D^{yy}_\kv &=& \frac{4 \pi \mu^2}{a^3} \left[\frac{1}{3} - \left(1-f_\kv\right) \sin^2 \theta_\kv\right], \label{eq:Dyy}\\
D^{zz}_\kv &=& \frac{4 \pi \mu^2}{a^3} \left[\frac{1}{3} - \left(1-f_\kv\right) \cos^2 \theta_\kv\right], \label{eq:Dzz}\\
D^{xy}_\kv &=& D^{yx}_\kv = 0 , \label{eq:Dxy}
\end{eqnarray}
\end{subequations}
with the form factor given in Eq.~(\ref{eq:formdef}).
Note that for $\kv = 0$ there is the general relation~\cite{Cohen95,Akhiezer68,Landau84,Syromyatnikov10}
\begin{equation}
D^{\alpha \beta}_{\kv = 0} = \frac{4 \pi \mu^2}{a^3} \left[\frac{1}{3} - \mathcal{N}_\alpha\right] \delta_{\alpha \beta},
\end{equation}
where $\mathcal{N}_\alpha$ is the geometry-dependent demagnetization factor. Eqs.~\eqref{eq:Dxx}--\eqref{eq:Dzz} are a special case of this relation.

\end{appendix}


\begin{thebibliography}{99}
%
\bibitem{Suhl57}
H. Suhl, {\it The theory of ferromagnetic resonance at high signal powers}, J. Phys. Chem. Solids {\bf{1}}, 209 (1957).
%
\bibitem{Schloemann60}
E. Schl\"{o}mann, J. J. Green, and U. Milano, 
{\it Recent Developments in Ferromagnetic Resonance at High Power Levels},
J. Appl. Phys. {\bf{31}}, 386S (1960); 
E. Schl\"{o}mann and R. I. Joseph, {\it Instability of Spin Waves and Magnetostatic Modes in a Microwave Magnetic Field Applied Parallel to the dc Field},
{\it{ibid.}} {\bf{32}}, 1006 (1961); 
E. Schl\"{o}mann and J. J. Green,
{\it Spin-Wave Growth Under Parallel Pumping}, {\it{ibid.}} {\bf{34}}, 1291 (1963).
%
\bibitem{Zakharov70}
V. E. Zakharov, V. S. L'vov, and S. S. Starobinets,
{\it Stationary nonlinear theory of parametric excitation of waves},
Zh. Eksp. Teor. Fiz. {\bf{59}}, 1200 (1970) 
[Sov. Phys. JETP {\bf{32}}, 656 (1971)].
%
\bibitem{Zakharov74}
V. E. Zakharov, V. S. L'vov, and S. S. Starobinets,
{\it Spin-wave turbulence beyond the parametric excitation threshold},
Usp. Fiz. Nauk {\bf{114}}, 609 (1974) [Sov. Phys.-Usp. {\bf{17}}, 896 (1975)].
%
\bibitem{Cherepanov93}
V. Charepanov, I. Kolokolov, and V. S. L'vov,
{\it The saga of YIG: spectra, thermodynamics, interaction and relaxation of magnons in a complex magnet},
Phys. Rept. {\bf{229}}, 81 (1993).
%
%
\bibitem{Araujo74}
C. B. Araujo, {\it Quantum-statistical
theory of the nonlinear excitation of magnons in parallel pumping experiments}, Phys. Rev. B {\bf{10}}, 3961 (1974).
%
\bibitem{Tsukernik75}
V. M. Tsukernik and R. P. Yankelevich,
{\it Stationary distribution of magnons following parametric excitation in ferromagnetic substance},
Zh. Eksp. Teor. Fiz. {\bf{68}}, 2116 (1975) [Sov. Phys. JETP {\bf{41}}, 1059 (1976)].
%
\bibitem{Vinikovetskii79}
I. A. Vinikovetskii, A. M. Frishman, and V. M. Tsukernik,
{\it Kinetic equation for a system of parametrically excited spin waves},
Zh. Eksp. Teor. Fiz. {\bf{76}}, 2110 (1979) [Sov. Phys. JETP {\bf{49}}, 1067 (1979)].
%
\bibitem{Lavrinenko81}
A. V. Lavrinenko, V. S. L'vov, G. A. Melkov, and V. B. 
Cherepanov, {\it "Kinetic" instability of a strongly nonequilibrium system of spin waves and tunable radiation of a ferrite}, Zh. Eksp. Teor. Fiz. {\bf{81}}, 1022 (1981) [Sov. Phys. JETP {\bf{54}}, 542 (1981)].
%
\bibitem{Zvyagin82} 
A. A. Zvyagin, V. Ya. Serebryannyi, A. M. Frishman, and V. M. Tsukernik, {\it Dynamics of spin waves under parametric excitation by a stepped periodic field of arbitrary amplitude}, Fiz. Nizk. Temp. \textbf{8}, 1205 (1982) [Sov. J. Low Temp. Phys. 8, 612 (1982)].
%
\bibitem{Zvyagin85}
A. A. Zvyagin and V. M. Tsukernik, 
{\it A change in equilibrium configuration of magnetic system during parametric excitation},
Fiz. Nizk. Temp. {\bf{11}}, 88 (1985) [Sov. J. Low Temp. Phys.
 {\bf{11}}, 47 (1985)].
%
\bibitem{Lim88}
S. P. Lim and D. L. Huber,
{\it Microscopic theory of spin-wave instabilities in parallel-pumped easy-plane ferromagnets},
Phys. Rev. B {\bf{37}}, 5426 (1988);
{\it Possible mechanism for limiting the number of modes in spin-wave instabilities in parallel pumping}, {\it{ibid.}} {\bf{41}}, 9283 (1990).
%
\bibitem{Kalafati89}
Yu. D. Kalafati and V. L. Safonov,
{\it Thermodynamic approach in the theory of paramagnetic resonance of magnons},
Zh. Eksp. Teor. Fiz. {\bf{95}}, 2009 (1989)
[Sov. Phys. JETP {\bf{68}}, 1162 (1989)].
%
\bibitem{Lvov94}
V. S. L'vov, {\it{Wave Turbulence Under Parametric Excitations}}, (Springer, Berlin, 1994).
%
\bibitem{Zvyagin07}
A. A. Zvyagin,
{\it Re-distribution (condensation) of magnons in a ferromagnet under pumping},
Fiz. Nizk. Temp. {\bf{33}}, 1248 (2007) [Sov. J. Low Temp. Phys.
 {\bf{33}}, 948 (2007)].
%
\bibitem{Rezende09}
S. M. Rezende, {\it Theory of microwave superradiance from a Bose-Einstein condensate of magnons}, Phys. Rev. B {\bf{79}}, 060410(R) (2009); {\it Theory of coherence in Bose-Einstein condensation phenomena in a microwave-driven interacting magnon gas}, {\it{ibid.}}, 174411 (2009).
%
\bibitem{Kloss10} 
T. Kloss, A. Kreisel, and P. Kopietz, {\it Parametric pumping and kinetics of magnons in dipolar ferromagnets}, Phys. Rev. B \textbf{81}, 104308 (2010).
%
\bibitem{Safonov13}
V. L. Safonov, {\it{Nonequilibrium Magnons}}, (Wiley-VCH, Weinheim, Germany, 2013).
%
\bibitem{Slobodianiuk17}
D. V. Slobodianiuk and O. V. Prokopenko,
{\it Kinetics of Strongly Nonequilibrium Magnon Gas Leading to Bose-Einstein Condensation},
J. Nano- Electron.  Phys. {\bf{9}}, 03033 (2017).
%
\bibitem{Demokritov06}
S. O. Demokritov, V. E. Demidov, O. Dzyapko, G. A. Melkov, A. A. Serga, B. Hillebrands, and A. N. Slavin,
{\it Bose-Einstein condensation of quasi-equilibrium magnons at room temperature under pumping}, Nature {\bf 443}, 430 (2006).
%
\bibitem{Demidov07}
V. E. Demidov, O. Dzyapko, S. O. Demokritov, G. A. Melkov, and A. N. Slavin,
{\it Thermalization of a Parametrically Driven Magnon Gas Leading to Bose-Einstein Condensation}, Phys. Rev. Lett. {\bf 99}, 037205 (2007).
%
\bibitem{Dzyapko07}
O. Dzyapko, V. E. Demidov, S. O. Demokritov, G. A. Melkov, and A. N. Slavin,
{\it Direct observation of Bose-Einstein condensation in a parametrically driven gas of magnons}, New J. Phys. {\bf 9}, 64 (2007).
%
\bibitem{Demidov08a}
V. E. Demidov, O. Dzyapko, S. O. Demokritov, G. A. Melkov, and A. N. Slavin,
{\it Observation of Spontaneous Coherence in Bose-Einstein Condensate of Magnons},
  Phys. Rev. Lett. {\bf 100}, 047205 (2008).
%
\bibitem{Demokritov08}
S. O. Demokritov, V. E. Demidov, O. Dzyapko, G. A. Melkov, and A. N. Slavin,
{\it Quantum coherence due to Bose–Einstein condensation of parametrically driven magnons},
  New J. Phys. {\bf 10}, 045029 (2008).
%
\bibitem{Demidov08b} V. E. Demidov, O. Dzyapko, M. Buchmeier, T. Stockhoff, G. Schmitz, G. A. Melkov, and S. O. Demokritov, {\it Magnon Kinetics and Bose-Einstein Condensation Studied in Phase Space}, Phys. Rev. Lett. {\bf 101}, 257201 (2008).
%
\bibitem{Serga14}
A. A. Serga, V. S. Tiberkevich, C. W. Sandweg, V. I. Vasyuchka, D. A. Bozhko, 
A. V. Chumak, T. Neumann, B. Obry, G. A. Melkov, A. N. Slavin, and 
B. Hillebrands, {\it Bose–Einstein condensation in an ultra-hot gas of pumped magnons}, Nat. Comm. {\bf{5}}, 3452 (2014).
%
\bibitem{Clausen15a}
P. Clausen, D. A. Bozhko, V. I. Vasyuchka, B. Hillebrands, G. A. Melkov, and A. A. Serga, {\it Stimulated thermalization of a parametrically driven magnon gas as a prerequisite for Bose-Einstein magnon condensation}, Phys. Rev. B {\bf 91}, 220402(R) (2015).
%
\bibitem{Clausen15b}
P. Clausen, D. A. Bozhko, V. I. Vasyuchka, G. A. Melkov, B. Hillebrands, and A. A. Serga,
{\it Supercurrent in a room-temperature Bose–Einstein magnon condensate},
Nature Physics {\bf{12}}, 1057 (2016).
%
\bibitem{Rueckriegel15}
A. R\"{u}ckriegel and P. Kopietz, {\it Rayleigh-Jeans Condensation of Pumped Magnons in Thin-Film Ferromagnets}, Phys. Rev. Lett. {\bf{115}}, 157203 (2015).
%
\bibitem{Holstein40} 
T. Holstein and H. Primakoff, {\it Field Dependence of the Intrinsic Domain Magnetization of a Ferromagnet}, Phys. Rev. \textbf{58}, 1098 (1940).
%
\bibitem{Rueckriegel12}
By quantizing the spin operators in a proper time-dependent reference frame, the magnon operators can always be defined such that
the expectation values $\langle a_{\bd{k}} ( t ) \rangle$ vanish identically, see
A. R\"{u}ckriegel, A. Kreisel,  and P. Kopietz, {\it Time-dependent spin-wave theory}, Phys. Rev. B {\bf{85}} , 054422 (2012). 
%
\bibitem{Fricke96}
J. Fricke, {\it{Transportgleichungen f\"{u}r quantenmechanische Vielteilchensysteme}}, 
(Cuvillier-Verlag, G\"{o}ttingen, 1996).
%
%
%
\bibitem{footnoteyig}
Very recently we have succeeded to solve the quantum kinetic equations for the
magnon distribution in YIG with a microscopically derived collision integral to explain
the effect of confluent magnon damping on the
parametric resonance of magnons in YIG reported by 
T. B. Noack, V. I. Vasyuchka, D. A. Bozhko, B. Heinz, P. Frey, D. V. Slobodianiuk, O. V. Prokopenko, G. A. Melkov, P. Kopietz, B. Hillebrands, and A. A. Serga,
{\it Enhancement of the Spin Pumping Effect by Magnon Confluence Process in YIG/Pt Bilayers},
Phys. Status Solidi B, \textbf{256}, 1900121 (2019).
%
\bibitem{Hick10}
J. Hick, F. Sauli, A. Kreisel, and P. Kopietz,
{\it Bose-Einstein condensation at finite momentum and magnon condensation in thin film ferromagnets},
Eur. Phys. J. B \textbf{78}, 429 (2010).
%
\bibitem{Rueckriegel14}
A. R\"{u}ckriegel, P. Kopietz, D. A. Bozhko, A. A. Serga, and B. Hillebrands,
{\it{Magnetoelastic modes and lifetime of magnons in thin yttrium iron garnet films}},
Phys. Rev. B {\bf{89}}, 184413 (2014).
%
\bibitem{Rezende06}
S. M. Rezende, F. M. de Aguiar, and A. Azevedo, {\it Magnon excitation by spin-polarized direct currents in magnetic nanostructures}, Phys. Rev. B \textbf{73}, 094402 (2006).
%
\bibitem{Kreisel09}
A. Kreisel, F. Sauli, L. Bartosch, and P. Kopietz,
{\it Microscopic spin-wave theory for yttrium-iron garnet films},
Eur. Phys. J. B  {\bf{71}}, 59 (2009). 
%
\bibitem{Tupitsyn08}
I. S. Tupitsyn, P. C. E. Stamp, and A. L. Burin,
{\it Stability of Bose-Einstein Condensates of Hot Magnons in Yttrium Iron Garnet Films},
Phys. Rev. Lett. {\bf{100}}, 257202 (2008).
%
\bibitem{Kalinkos86}
B. A. Kalinikos, and A. N. Slavin,
{\it Theory of dipole-exchange spin wave spectrum for ferromagnetic films with mixed exchange boundary conditions},
J. Phys. C {\bf{19}}, 7013 (1986).
%
\bibitem{Filho00} 
R. N. Costa Filho, M.G. Cottam, and G.A. Farias, {\it Microscopic theory of dipole-exchange spin waves in ferromagnetic films: Linear and nonlinear processes}, Phys. Rev. B \textbf{62}, 6545 (2000).
%
\bibitem{Cohen95} 
M. H. Cohen and F. Keffer, {\it Dipolar Sums in the Primitive Cubic Lattices}, Phys. Rev. \textbf{99}, 1128 (1955).
%
\bibitem{Akhiezer68}
A. I. Akhiezer, V. G. Bar'yakhtar, and S. V. Peletminskii, \textit{Spin Waves} (North-Holland, Amsterdam, 1968).
%
\bibitem{Landau84} 
L. D. Landau and E. M. Lifshitz, \textit{Electrodynamics of Continuous Media} (Pergamon, Oxford, 1984).
%
\bibitem{Syromyatnikov10}
A. V. Syromyatnikov, {\it Anomalously large damping of long-wavelength quasiparticles caused by long-range interaction}, Phys. Rev. B \textbf{82}, 024432 (2010).
%
\end{thebibliography}
\end{document}